\documentclass[journal]{IEEEtran}

\usepackage{amsmath}
\usepackage{amsfonts}
\usepackage{amssymb}
\usepackage{amsthm}
\usepackage{tabularx}
\usepackage{mathrsfs} 

\usepackage{url}
\usepackage[colorlinks]{hyperref}

\newtheorem{remark}{Remark}

\usepackage{stfloats}
\usepackage{float}
\usepackage{graphicx}
\usepackage{xcolor}
\usepackage{subfigure}

\makeatletter
\def\blfootnote{\xdef\@thefnmark{}\@footnotetext}
\makeatother

\begin{document}
	
\title{Enormous Fluid Antenna Systems (E-FAS): Channel Estimation} %with Imperfect CSI }

\author{Farshad~Rostami~Ghadi,~\IEEEmembership{Senior~Member},~\textit{IEEE}, 
	    Kai-Kit~Wong,~\IEEEmembership{Fellow},~\textit{IEEE},\\
     	    Masoud~Kaveh,~\IEEEmembership{Member},~\textit{IEEE}, 
	    Hao~Xu,~\IEEEmembership{Senior~Member},~\textit{IEEE},
            Baiyang Liu,~\IEEEmembership{Senior Member,~IEEE}, 
	    Kin-Fai~Tong,~\IEEEmembership{Fellow},~\textit{IEEE}, and 
            Chan-Byoung Chae,~\IEEEmembership{Fellow,~IEEE}
\vspace{-7mm}
}

\maketitle
%\blfootnote{\noindent Copyright (c) 2015 IEEE. Personal use of this material is permitted. However, permission to use this material for any other purposes must be obtained from the IEEE by sending a request to pubs-permissions@ieee.org.} 

\blfootnote{The work of  F. Rostami Ghadi and K. K. Wong is supported by the Engineering and Physical Sciences Research Council (EPSRC) under Grant EP/W026813/1. The work of H. Xu is supported by National Natural Science Foundation of China (NSFC) under Grants 62501152 and U25A20398, and the Fundamental Research Funds for the Central Universities under grant 2242025R10001. The work of B. Liu and K.-F. Tong is supported by Hong Kong Metropolitan University, Staff Research Start-Up Fund under Grant FRSF/2024/03. The work of C.-B. Chae is supported by the Institute for Information and Communication Technology Planning and Evaluation (IITP)/National Research Foundation of Korea (NRF) grant funded by the Ministry of Science and ICT (MSIT), South Korea, under Grant RS-2024-00428780 and 2022R1A5A1027646.}

\blfootnote{\noindent F. Rostami Ghadi and K. K. Wong are with the Department of Electronic and Electrical Engineering, University College London, London, WC1E 7JE, United Kingdom. K. K. Wong is also affiliated with Yonsei Frontier Lab, Yonsei University, Republic of Korea (e-mail: $\rm \{f.rostamighadi, kai\text{-}kit.wong\}@ucl.ac.uk$).}
\blfootnote{\noindent M. Kaveh is with the Department of Information and Communication Engineering, Aalto University, Espoo, Finland (e-mail: $\rm masoud.kaveh@aalto.fi$).}
\blfootnote{\noindent H. Xu is with the National Mobile Communications Research Laboratory, Southeast University, Nanjing 210096, China (e-mail: $\rm hao.xu@seu.edu.cn$).}
\blfootnote{\noindent B. Liu and K. F. Tong are with the School of Science and Technology, Hong Kong Metropolitan University, Hong Kong SAR, China. B. Liu is also affiliated with School of Artificial Intelligence, Shenzhen Technology University, Shenzhen, China (e-mail: $\rm \{byliu, ktong\}@hkmu.edu.hk$).}
\blfootnote{\noindent C.-B. Chae is with the School of Integrated Technology, Yonsei University, Seoul, 03722, Republic of Korea (e-mail: $\rm cbchae@yonsei.ac.kr$).}

\blfootnote{\noindent \em Corresponding Author: Kai-Kit Wong.}
%\IEEEpeerreviewmaketitle

\begin{abstract}
Enormous fluid antenna systems (E-FAS) have recently emerged as a new wireless architecture in which intelligent metasurfaces act as guided electromagnetic interfaces, enabling surface wave (SW) propagation with much lower attenuation and more control than conventional space wave transmission. While prior work has reported considerable power gains under perfect channel state information (CSI), the impact of practical channel acquisition on E-FAS performance remains largely unexplored. In this paper, we present the first comprehensive analysis of E-FAS-assisted downlink transmission under pilot-based channel estimation. We develop an estimation framework for the equivalent end-to-end channel and derive closed-form expressions for the statistics of the minimum mean square error (MMSE) channel estimate and its estimation error. Building on these results, we analyze both single-user and multiuser operation while explicitly accounting for the training overhead. {\color{blue}For the single-user case, we characterize an outage probability upper bound and an achievable rate lower bound under imperfect CSI and reveal a fixed training high-signal-to-noise ratio (SNR) effective signal-to-interference-plus-noise ratio (SINR) saturation caused by residual self-interference.
	%For the single-user case, we characterise the outage probability and achievable rate with imperfect CSI, and reveal an inherent signal-to-noise ratio (SNR) saturation phenomenon caused by residual self-interference. 
For the multiuser case, we study zeroforcing (ZF) precoding based on imperfect channel estimates and illustrate that under fixed pilot resources, residual inter-user interference produces an interference-limited high-SNR regime.} %For the multiuser case, we study zero-forcing (ZF) precoding based on imperfect channel estimates and show that the system becomes interference-limited in the high SNR regime because of residual inter-user interference. 
Also, we quantify the trade-off between spatial multiplexing gains and pilot overhead when the number of users increases. Analytical findings are validated via Monte Carlo simulations and benchmarked against least-squares (LS) estimation and conventional non-E-FAS transmission. %The results reveal that despite CSI imperfections and training costs, E-FAS retains substantial performance advantages and provides robustness enabled by its amplified large-scale channel gain.
{\color{blue}The results show that a normalized E-FAS routing gain enhancement improves both the received data signal power and the effective pilot quality despite CSI imperfections and training overhead.}
\end{abstract}


\vspace{-2mm}
\section{Introduction}\label{sec:intro}
\subsection{Background}
{\color{blue}\IEEEPARstart{T}{he increasing} demand for ultra-reliable super-high rate wireless connectivity has motivated the use of higher frequency bands in next-generation systems \cite{matt20216g}. Nevertheless, space wave propagation at high frequencies suffers from severe spreading loss, blockage, and limited diffraction capability \cite{tariq2020spec,rappa2017over}. Reconfigurable intelligent surfaces (RISs) can partially reshape the propagation environment via smart reflection, but remain subject to space wave propagation, cascaded path loss, and potentially large channel acquisition overhead \cite{basar2019wire,huang2019rec}. 

Beyond manipulating the propagation environment through RISs, the emerging concept of fluid antenna systems (FAS) focuses upon enhancing spatial adaptability at the transceiver side and has become a new communication paradigm \cite{wang2021fluid,wong2020perf,new2025tut,New-jsac2026,hong2025contemporary,Hong-fama2026}. Recent hardware developments have demonstrated several FAS implementations \cite{shen2024design,zhang2024pixel,Wong-wc2026,tong2025design,Shamim-2025}, while representative analytical studies have characterized their diversity and spatial correlation properties \cite{ghadicopula,khammassi2023new,ramirez2024new}. FAS emphasizes on the antenna reconfigurability for shape and position flexibility for diversity and multiplexing benefits. Even on one single radio frequency (RF) chain, FAS can dynamically reconfigure its radiation position within a confined spatial region, effectively exploiting spatial diversity without increasing hardware complexity much. This reconfigurability enables improved link reliability and diversity gains, particularly in small form factor devices.  FAS concepts have also been combined with RIS \cite{ghadi2024perf,11185049} and extended to fluid RIS (FRIS) \cite{salem2025first,xiao2025fluid,ghadi2025perfo,xiao2026fluid,rostami2025perf}, and fluid  integrated reflecting and emitting surfaces (FIRES) \cite{Ghadi2025fires}. 

%These architectures improve spatial reconfigurability but do not directly address the guided surface wave (SW) routing mechanism considered in this work.

Nevertheless, conventional FAS improves transceiver side spatial adaptability by reconfiguring antenna positions or radiation states within a confined local aperture. Its benefit is therefore limited by the available aperture and local channel diversity. More importantly, it does not modify the end-to-end propagation route and cannot directly eliminate spherical spreading, blockage, or severe high frequency propagation loss. RIS- and FRIS-assisted links also predominantly manipulate space wave propagation and generally retain cascaded propagation and channel acquisition burdens. These limitations motivate a propagation infrastructure solution rather than only local antenna or reflection reconfiguration.}

\vspace{-2mm}
\subsection{A New Communication Paradigm and Motivation}
Recognizing that the propagation characteristics constrained by space-wave propagation mechanisms and 3D power spreading are the main problem, the spirit of position reconfigurability in FAS could indeed be useful, if it is applied in a much larger scale. This has led to the concept of enormous FAS (E-FAS) \cite{Wong-2021swc,wong2025enor,rostami2026enorm}. Specifically, an E-FAS is an ensemble of one or more coordinated intelligent surfaces (or engineered metasurfaces) that helps establish communication between the transmitter(s) and receiver(s). A distinctive feature is that the engineered metasurfaces can serve as guided electromagnetic interfaces instead of reflectors in conventional RISs. In particular, incident space waves are converted into surface waves (SWs) that propagate along quasi-two-dimensional structures with cylindrical spreading before being re-radiated towards intended users. This mechanism significantly reduces attenuation compared to spherical space wave spreading and enables energy routing along walls, ceilings, and building facades. As a result, E-FAS provides substantial large-scale power saving while preserving simplicity in end-to-end optimization.\footnote{\color{blue}It is also useful to distinguish E-FAS from the recently introduced scalable FAS (SFAS) \cite{wu2026sfas}. SFAS dynamically changes the physical aperture configuration at the transceiver, thereby creating configuration array manifolds and observation geometries for array signal processing. E-FAS instead reconfigures the propagation infrastructure through distributed engineered surfaces that convert incident space waves into guided SWs, route electromagnetic energy along programmable pathways, and re-radiate it toward the intended users. Thus, SFAS primarily modifies the transceiver observation aperture, whereas E-FAS modifies the end-to-end propagation route.}

{\color{blue}From an implementation perspective, E-FAS can be constructed from interconnected programmable metasurface tiles incorporating SW launchers, guided transmission sections, electronically controlled routing junctions, and re-radiating apertures. These building blocks are broadly compatible with current printed metasurface and reconfigurable antenna fabrication technologies \cite{Liu-efas2026}. Important practical challenges include route-based attenuation, launcher and re-radiation efficiencies, mutual coupling, channel calibration, and distributed routing control. This work's aim is to develop a physically grounded analytical framework whose parameters can be obtained from measurement or electromagnetic simulation.}

%Despite the initial vision first proposed in 2021 \cite{Wong-2021swc}, little is known regarding the performance of E-FAS and its design aspects. In \cite{wong2025enor}, the electromagnetic modelling of E-FAS was discussed while \cite{rostami2026enorm} presented initial evidence of promising improvements in outage probability and ergodic rate. However, like most physical-layer techniques, CSI is crucial for E-FAS but this issue remains unexplored. In particular, the equivalent end-to-end channel comprises a slowly varying SW routing gain and a short-term fading component, while the underlying multi-segment propagation and hidden SW routing stages are not directly observable at the base station (BS). These unique features fundamentally differentiate E-FAS from conventional space wave multiple-input multiple-output (MIMO) and call for an explicit investigation of pilot-based channel estimation and the resulting performance trade-offs under realistic CSI estimation techniques and training overhead.

{\color{blue}Despite the initial E-FAS vision introduced in \cite{Wong-2021swc}, only limited results are currently available regarding its communication performance and practical design. Electromagnetic aspects of E-FAS were discussed in \cite{wong2025enor}, while the initial performance analysis in \cite{rostami2026enorm} demonstrated substantial outage and rate gains under perfect channel state information (CSI). The latter assumption, however, may considerably overestimate the practically achievable performance. For example, under the representative system configuration considered in Section~\ref{sec:num}, neglecting channel estimation errors overestimates the single-user achievable rate bound by approximately $26.4\%$ and $39.8\%$ at downlink signal-to-noise ratios (SNRs) of $20$ and $30$ dB, respectively. For multiuser zeroforcing (ZF) transmission, the corresponding sum-rate overestimations are approximately $33.9\%$ and $48.5\%$. These losses become increasingly pronounced at high downlink SNR when the pilot resources are fixed, demonstrating that channel acquisition is a first-order design consideration for E-FAS-assisted transmission.}

\vspace{-2mm}
\subsection{Key Questions}
{\color{blue}An E-FAS channel consists of a slowly varying topology dependent routing gain and a fast fading component. Hence, the end-to-end channel has a two-timescale structure and must be inferred through pilot-based training over finite coherence intervals.} Imperfect CSI is thus inevitable and degrades system performance. More importantly, imperfect CSI introduces two critical phenomena that are absent under perfect CSI: (i) residual self-interference, referring to the non-vanishing interference component induced by channel estimation errors due to beamforming mismatch, leading to SNR saturation at high transmit powers in single-user transmission, and (ii) residual multiuser interference under ZF precoding, resulting in the interference-limited regime at high SNR. {\color{blue}At low downlink power, thermal noise dominates the system performance. As the transmit power increases, however, the residual interference caused by imperfect channel estimates becomes dominant unless the pilot resources are increased sufficiently.} Whether the substantial power gains by E-FAS persist under realistic CSI acquisition constraints thus remains an open question.

{\color{blue}
Recent channel acquisition studies address several different estimation objectives. FAS methods typically reconstruct the spatial channel from observations at a subset of antenna positions \cite{new2025channel,zhang2025succ,tang2025accurate}, whereas RIS methods estimate the high-dimensional cascaded base station (BS)-RIS-user channel \cite{wei2021channel}. Additionally, the fluid antenna-enabled hybrid analog-digital (FA-HAD) architecture in \cite{tian2026hybrid} instead considers two-dimensional direction-of-arrival estimation and spatial covariance reconstruction. These approaches are not directly comparable with the present framework because they estimate different physical quantities under different hardware and observation models. Unlike conventional FAS channel reconstruction over a local reconfigurable aperture with a fixed propagation route, in this paper, our aim is to estimate the composite E-FAS channel whose topology-dependent routing gain and fast fading evolve over different timescales. Consequently, existing FAS and RIS methods cannot be applied directly without reformulating the observation model and estimation target.}

The above observations  motivate us for a systematic investigation of E-FAS-assisted transmission under imperfect CSI. In this regard, several fundamental questions arise:
\begin{enumerate}
\item How does pilot-based estimation affect the equivalent end-to-end channel statistics in E-FAS-assisted systems?
\item Also, does the SW-induced large-scale gain translate into proportional performance gains under imperfect CSI?
\item How do training overhead and spatial multiplexing interact in multiuser E-FAS-assisted systems?

%\item Under what conditions does the system become noise-limited or interference-limited?
\item {\color{blue}How does imperfect CSI govern the transition from a noise-limited to a residual interference-limited regime, and under what pilot resource scaling can the resulting high-SNR ceiling be avoided?}
\end{enumerate}

Answering the above is essential to understand the robustness of E-FAS and to quantify the trade-offs between SW gain, estimation accuracy, and spatial degrees of freedom (DoF).

\vspace{-2mm}
\subsection{Contributions}
{\color{blue}This paper develops a two-timescale channel acquisition and performance analysis framework for E-FAS-assisted downlink
transmission under imperfect CSI. The main contributions of this paper are summarized as follows:}
\begin{itemize}
%\item We develop a pilot channel estimation framework for the end-to-end channel and obtain closed-form expressions for the variances of the minimum mean square error (MMSE) estimate and the associated estimation error.
{\color{blue}\item First, we formulate E-FAS channel acquisition as a two-timescale estimation problem. The topology-dependent routing gain is inferred by long-term pilot energy averaging, while the fast-fading component is estimated using minimum mean square error (MMSE) in each coherence block. This formulation explicitly captures the coupling between SW routing and effective pilot quality.
%\item For the single-user case, we derive analytical expressions for the outage probability and the achievable rate under imperfect CSI while explicitly accounting for the training overhead. We characterize the resulting SNR saturation behavior and quantify the loss relative to perfect CSI.
\item For single-user maximum ratio transmission (MRT), we derive expressions for an outage probability upper bound and an achievable rate lower bound with training overhead and characterize how the routing gain and estimation error jointly determine the fixed training high-SNR rate ceiling and the loss relative to perfect CSI.
%\item For the multi-user downlink, we analyze ZF precoding based on imperfect channel estimates and obtain tractable expressions for the residual interference-limited signal-to-interference-plus-noise ratio (SINR) and ergodic sum-rate. We reveal how the system transitions from noise-limited to interference-limited regimes at high SNR.
%\item For multi-user ZF transmission, we obtain tractable signal-to-interference-plus-noise ratio (SINR) and ergodic sum-rate expressions under imperfect CSI and characterize the effects of routing gain, residual interference, pilot overhead, and spatial multiplexing on the transition from the noise-limited to the interference-limited regime.
\item Moreover, for multiuser ZF transmission, we characterize how the E-FAS routing gain, imperfect CSI, pilot overhead, and spatial multiplexing jointly govern residual interference, high-SNR saturation, and the transition from noise-limited to interference-limited operation.}
{\color{blue}\item We derive the exact finite sample distribution and variance of the slow timescale routing gain estimator and quantify the excess channel estimation mean-square-error (MSE) caused by mismatched routing gain knowledge. Based on these results, we obtain an explicit minimum window selection criterion that satisfies a prescribed estimation accuracy requirement while respecting the routing stationarity interval.}
%\item Also, we investigate the interplay between pilot length, coherence interval, and the number of users, exposing the trade-off between spatial multiplexing gain and training overhead in E-FAS-assisted systems.
{\color{blue}\item Furthermore, we formulate a joint pilot-power and pilot-length allocation problem under pilot-energy and peak-power constraints. By exploiting the monotonic dependence of the estimation quality on the accumulated pilot energy, we reduce the mixed-integer problem to a one-dimensional ($1$-D) search that yields the optimal resource allocation for both single-user and multiuser operations.
%\item Extensive Monte Carlo simulations validate the accuracy of the derived analytical expressions and confirm the tight agreement between theory and simulation. MMSE and least squares (LS) estimators are benchmarked against the perfect CSI case and conventional non-E-FAS transmission. The results show that MMSE consistently outperforms LS, particularly at low pilot SNR and short training lengths, and that E-FAS maintains a pronounced SNR and sum-rate advantage over conventional transmission even under the practical situation where CSI is imperfect.
\item Monte Carlo simulations validate the analytical expressions and quantify the effects of routing gain, finite window uncertainty, pilot resource allocation, and imperfect CSI. MMSE and least squares (LS) estimation are compared with perfect CSI and conventional non-E-FAS benchmarks under common channel and pilot settings. Additional simulations assess robustness to spatial correlation, unequal routing gains, and pilot contamination.}
\end{itemize}

\vspace{-2mm}
\subsection{Organization and Notations}
The remainder of this paper is organized as follows. Section~\ref{sec:system} describes the system and channel models for E-FAS. Section~\ref{sec-pilot} presents the pilot-based channel estimation framework and derives the statistics of the channel estimate and its error. Section~\ref{sys-single} analyzes the single-user case while Section~\ref{sec:mult} investigates multiuser ZF precoding with imperfect CSI estimates. Numerical results are provided in Section~\ref{sec:num}, and finally, we conclude this paper in Section~\ref{sec:conc}.

\textit{Notations}---Throughout this paper, scalars are denoted by italic letters, vectors by bold lowercase letters, and matrices by bold uppercase letters. The operators $(\cdot)^T$ and $(\cdot)^H$ denote the transpose and Hermitian transpose, respectively. The Euclidean norm of a vector is denoted by $\|\cdot\|$, and $|\cdot|$ represents the absolute value of a scalar. The expectation operator is denoted by $\mathbb{E}\{\cdot\}$. The notation $\mathcal{CN}(\mathbf{0},\mathbf{R})$ denotes a circularly symmetric complex Gaussian distribution with zero mean and covariance matrix $\mathbf{R}$. The identity matrix of dimension $M$ is denoted by $\mathbf{I}_M$. The Gamma function is denoted by $\Gamma(\cdot)$, and the lower incomplete Gamma function by $\Upsilon(\cdot,\cdot)$. Logarithms are taken with base $2$ unless otherwise specified.

\begin{figure}[]
\centering
\includegraphics[width=.65\columnwidth]{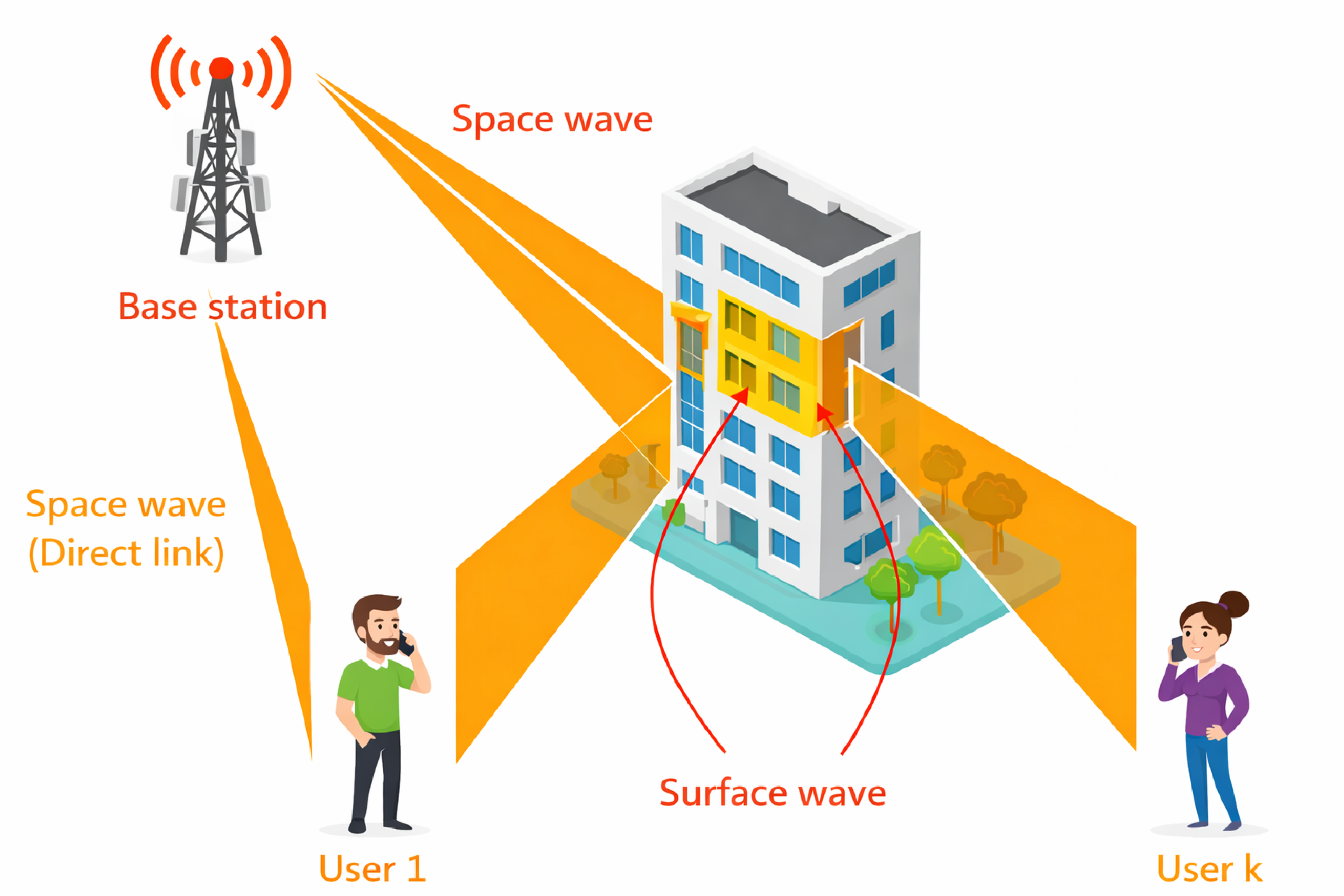}
\caption{Illustration of the E-FAS-assisted downlink system.}\label{fig:model}
\vspace{-4mm}
\end{figure}

\vspace{-4mm}
\section{System and Channel Model}\label{sec:system}
\subsection{System Architecture}
As illustrated in Fig.~\ref{fig:model}, we consider the downlink of an E-FAS-assisted wireless network, in which an $M$-antenna BS simultaneously serves $K$ single-antenna users. The propagation environment is instrumented with an E-FAS comprising distributed programmable metasurface tiles deployed on surrounding structures such as walls, ceilings, or building facades. Unlike conventional reconfigurable metasurfaces that primarily reflect space waves, the E-FAS converts incident space waves into guided SWs, routes electromagnetic energy along engineered surface pathways, and re-radiates the signal toward intended users through programmable launcher structures. 

For the model considered, we let $x_u \in \mathbb{C}$ denote the information symbol intended for user $u$, satisfying $\mathbb{E}\{|x_u|^2\}=1$ and the BS applies linear precoding and transmits
\begin{align}
\mathbf{x} = \sum_{u=1}^{K} \sqrt{P_u}\,\mathbf{w}_u x_u,
\end{align}
where $\mathbf{w}_u \in \mathbb{C}^{M}$ is the precoding vector associated with user $u$, $P_u$ is the allocated transmit power, and $\sum_{u=1}^{K} P_u = P$ denotes the total BS transmit power. Therefore, the received signal at user $u$ is then given by
\begin{align}
y_u = \mathbf{h}_{u}^{H}\mathbf{x} + n_u,
\end{align}
where $\mathbf{h}_u \in \mathbb{C}^{M}$ represents the equivalent end-to-end BS-user channel induced by E-FAS-assisted propagation, and $n_u \sim \mathcal{CN}(0,\sigma^2)$ is additive white Gaussian noise (AWGN).

\vspace{-2mm}
{\color{blue}
\vspace{-2mm}
{\color{blue}\subsection{Physics-Based Two-Timescale Channel Representation}}

The E-FAS propagation mechanism comprises multiple physical segments, including space wave transmission from the BS to the surface, conversion to guided SWs, propagation along programmable surface routes, and re-radiation toward the users. The routing configuration and launcher arrangement generally vary more slowly than the small-scale fading associated with the wireless segments. This motivates a two-timescale channel representation. Let $\mathcal P_u$ denote the active SW route for user $u$, comprising $L_u$ surface segments and $J_u$ junctions or bends. A physically grounded representation of its average end-to-end power gain is given by
%\vspace{-2mm}
\begin{align}
	\beta_u(\mathcal P_u)
	={}&
	\ell_{{\rm B}\rightarrow{\rm in},u}
	\!\left(d_{{\rm B}\rightarrow{\rm in},u}\right)
	\eta_{{\rm in},u}
	\prod_{\ell=1}^{L_u}
	\exp\!\left(-\alpha_{u,\ell}d_{u,\ell}\right)
	\nonumber\\
	&\times
	\prod_{j=1}^{J_u}\xi_{u,j}\,
	\eta_{{\rm out},u}
	\ell_{{\rm out}\rightarrow u}
	\!\left(d_{{\rm out}\rightarrow u}\right),
	\label{eq:physical_routing_gain}
\end{align}
where the two $\ell(\cdot)$ terms denote the average space wave power gains before and after SW propagation, $\eta_{{\rm in},u}$ and $\eta_{{\rm out},u}$ are the input coupling and re-radiation efficiencies, respectively, $\alpha_{u,\ell}$ is the power attenuation coefficient of surface segment $\ell$, and $\xi_{u,j}\in(0,1]$ represents the power transmission efficiency of junction or bend $j$. Hence, the route geometry affects $\beta_u$ through the selected segments, their lengths, and the number and type of junctions. For example, a far-field space wave segment may be represented as $\ell(d)=Cd^{-\nu}$, while measured or electromagnetic path gain models can be substituted without altering the subsequent estimation framework.

For notational simplicity, we write $\beta_u\equiv\beta_u(\mathcal P_u)$. Although the individual factors in \eqref{eq:physical_routing_gain} may not be separately observable at the BS, their aggregate effect is inferred from the received pilot observations. To distinguish this average power gain from the spatial structure of the channel, we adopt
%Let $\beta_u>0$ denote the average end-to-end power gain associated with the current E-FAS routing configuration. It aggregates the space wave path losses, launcher and re-radiation efficiencies, SW attenuation, routing distance, bends, junction losses,
%and other topology-dependent propagation effects. To distinguish
%this average power gain from the spatial structure of the channel,
%we adopt the generalized representation
\begin{align}
\mathbf h_u=\sqrt{\beta_u}\left(\overline{\mathbf h}_u+\mathbf R_u^{1/2}\mathbf g_u\right),~\mbox{where }\mathbf g_u\sim\mathcal{CN}(\mathbf 0,\mathbf I_M),\label{eq:general_channel}
\end{align}
where $\overline{\mathbf h}_u$ represents a deterministic specular component and $\mathbf R_u\succeq\mathbf 0$ is the spatial covariance matrix of the diffused component. We normalize these quantities such that
\begin{align}
\left\|\overline{\mathbf h}_u\right\|^2+\operatorname{tr}(\mathbf R_u)=M,
\end{align}
so that $\mathbb E\{\|\mathbf h_u\|^2\}=M\beta_u$. Thus, 
$
	\mathbf h_u
	\sim
	\mathcal{CN}\left(
	\sqrt{\beta_u}\overline{\mathbf h}_u,
	\beta_u\mathbf R_u
	\right)$. The vector $\overline{\mathbf h}_u$ represents a dominant line-of-sight (LoS) or spherical wave near-field response, whereas $\mathbf R_u$ can capture spatial correlation induced by the propagation geometry, active re-radiating elements, common scattering clusters, and mutual coupling. The dependence of $\beta_u$, $\overline{\mathbf h}_u$, and $\mathbf R_u$ on the E-FAS routing configuration is kept implicit for notational simplicity. The closed-form analysis in the following sections focuses on the rich scattering special case $\overline{\mathbf h}_u=\mathbf 0$ and $\mathbf R_u=\mathbf I_M$, so
\begin{equation}
	\mathbf h_u=\sqrt{\beta_u}\mathbf g_u\sim
	\mathcal{CN}(\mathbf 0,\beta_u\mathbf I_M).
	\label{eq:channel}
\end{equation}
This specialization isolates the impact of topology-dependent routing-gain acquisition while retaining analytical tractability. In particular, the routing gain $\beta_u$ evolves over the slow routing stationarity interval $T_r$, whereas $\mathbf g_u$ varies over the channel coherence interval $T_c$, with $T_r\gg T_c$.}

\vspace{3mm}
\subsection{Signal Model}
Substituting the equivalent channel model into the received signal expression yields
\begin{align}
y_u = \sqrt{\beta_u}\,\mathbf{g}_u^{H}\mathbf{x} + n_u.
\end{align}
In the single-user case, i.e., $K=1$, the received signal is simplified to
\begin{align}
y = \sqrt{P\beta}\,\mathbf{g}^{H}\mathbf{w} x + n,
\end{align}
where $\mathbf{w}$ is a unit-norm precoding vector. 

In the multiuser case, i.e., $K \geq 2$, substituting the transmit signal into the received signal expression yields
\begin{align}
y_u = \sqrt{P_u \beta_u}\,\mathbf{g}_u^{H}\mathbf{w}_u x_u
+ \sum_{i \neq u} \sqrt{P_i \beta_u}\,\mathbf{g}_u^{H}\mathbf{w}_i x_i,
+ n_u,\label{eq:y_u}
\end{align}
where the first term corresponds to the desired signal component and the second term represents inter-user interference.

Note that the BS does not observe the individual physical segments, e.g., SW routing or launcher parameters, separately. Instead, it acquires only the composite channel vector $\mathbf{h}_u$. Furthermore, due to the two-timescale nature of E-FAS-assisted propagation, the large-scale routing gain $\beta_u$ and the small-scale fading component $\mathbf{g}_u$ may require different estimation strategies, which will be developed next.

\vspace{-2mm}
\section{Two-Timescale Channel Estimation Framework}\label{sec-pilot}
In this section, we develop a channel estimation framework tailored to the intrinsic two-timescale structure of E-FAS-assisted propagation introduced in Section~\ref{sec:system}. In particular, the equivalent end-to-end channel comprises a slowly varying topology-dependent routing gain and a fast small-scale fading component.  This separation naturally motivates a decoupled estimation procedure: (i) the routing gain is learned over a slow timescale via long-term pilot energy averaging, while (ii) the fast fading component is estimated per coherence block using MMSE principles. Unlike conventional multiple-input multiple-output (MIMO) settings where large-scale fading is purely geometry-induced and not reconfigurable, in E-FAS, the routing gain depends on the SW topology and directly affects the pilot observation quality, thereby coupling channel acquisition with routing design. {\color{blue}Conditioned on known routing gain and spatial statistics, the per-block MMSE estimator actually has the conventional Bayesian form. The distinction from conventional MIMO and FAS estimation lies in acquiring the unknown topology-dependent routing gain over a slow timescale and incorporating its finite window uncertainty into the fast timescale composite channel estimate.}

\vspace{-2mm}
\subsection{Training Protocol}
Consider a time-division duplex (TDD) system and exploit channel reciprocity for downlink channel acquisition via uplink pilot transmission. During each coherence block of length $T_c$, the users transmit pilot sequences of length $\tau_p$ symbols.  Let $\rho_p$ denote the pilot transmit power per symbol. Each user $u$ is assigned an orthogonal pilot sequence $\boldsymbol{\phi}_u \in \mathbb{C}^{\tau_p}$ of length $\tau_p$, satisfying $\boldsymbol{\phi}_u^{H}\boldsymbol{\phi}_i = 0$ for $u \neq i$ and $\|\boldsymbol{\phi}_u\|^2 = \tau_p$.  During the uplink training phase, the BS correlates the received pilot signal with $\boldsymbol{\phi}_u$ to isolate the contribution of user $u$. Then the resulting sufficient statistic is given by \cite{ngo2013energy}
\begin{align}
\mathbf{y}_{p,u}=\sqrt{\tau_p \rho_p}\,\mathbf{h}_u+\mathbf{n}_{p,u},
\end{align}
where $\mathbf{n}_{p,u} \sim \mathcal{CN}(\mathbf{0}, \sigma^2 \mathbf{I}_M)$ is AWGN after despreading.

Now, recalling from \eqref{eq:channel}, the observation becomes
\begin{align}
\mathbf{y}_{p,u}=\sqrt{\tau_p \rho_p \beta_u}\,\mathbf{g}_u+\mathbf{n}_{p,u}.\label{eq:chan}
\end{align}
Importantly, the BS observes only the composite end-to-end channel realization and does not have direct access to the routing gain $\beta_u$ or the individual physical propagation segments.  Nevertheless, under physically consistent modeling of the multi-segment propagation, the equivalent channel admits a tractable Gaussian representation in which the topology-dependent routing gain appears as a distinct large-scale parameter evolving on a slow timescale.  This structure preserves analytical tractability while explicitly coupling channel acquisition with routing configuration.

{\color{blue}\begin{remark}[Nonorthogonal Pilots and Pilot Contamination]
The orthogonal pilot model is feasible in the considered setting when $\tau_p\geq K$. With pilot reuse or $\tau_p<K$, define
	$
			c_{ui}
			\triangleq
			\frac{\boldsymbol{\phi}_u^{H}\boldsymbol{\phi}_i}{\tau_p}$,  $c_{uu}=1
	$, 
		and $q_i=\tau_p\rho_{p,i}$. Under the zero-mean correlated
		channel model, the despread observation becomes
		\begin{align}
			\mathbf y_{p,u}
			=
			\sum_{i=1}^{K}
			\sqrt{q_i}\,c_{ui}\mathbf h_i
			+
			\mathbf n_{p,u}.
			\label{eq:nonorthogonal_pilot_observation}
		\end{align}
		Defining
	$
			\boldsymbol{\Psi}_{p,u}
			=
			\sum_{i=1}^{K}
			q_i|c_{ui}|^2\beta_i\mathbf R_i
			+
			\sigma^2\mathbf I_M
	$, 
		the corresponding linear MMSE (LMMSE) estimate and error covariance are
		\begin{align}
			\widehat{\mathbf h}_u
			&=
			\sqrt{q_u}\beta_u\mathbf R_u
			\boldsymbol{\Psi}_{p,u}^{-1}
			\mathbf y_{p,u},
			\\
			\mathbf C_{\widetilde h,u}
			&=
			\beta_u\mathbf R_u
			-
			q_u\beta_u^2\mathbf R_u
			\boldsymbol{\Psi}_{p,u}^{-1}\mathbf R_u.
		\end{align}
		The orthogonal pilot expressions follow by setting
		$c_{ui}=0$ for $i\neq u$. With pilot reuse, the interference terms
		remain in $\boldsymbol{\Psi}_{p,u}$ and the estimation errors of
		different users are generally coupled. Therefore, the closed-form ZF rate analysis  specializes to the orthogonal pilot case.
\end{remark}}

\vspace{-3mm}
\subsection{Estimation of the Short-Term Fading Component}
%Since the slow routing gain $\beta_u$ evolves on a much larger timescale than the coherence block, it can be treated as quasi-static during fast fading estimation. Conditioned on this slowly varying routing parameter, the estimation problem reduces to recovering the small-scale fading vector $\mathbf g_u$ from the sufficient statistic $\mathbf y_{p,u}$.  Although the resulting estimator follows standard MMSE principles for Gaussian channels, its performance is directly governed by the topology-dependent routing gain $\beta_u$, which determines the effective pilot observation quality. This coupling between routing configuration and estimation accuracy is intrinsic to E-FAS-assisted propagation.

{\color{blue}Since the routing gain and the spatial channel statistics evolve more slowly than the instantaneous fading, we condition on $\beta_u$, $\overline{\mathbf h}_u$, and $\mathbf R_u$ during each coherence block. Let $q\triangleq\tau_p\rho_p$. For the general model in \eqref{eq:general_channel}, the MMSE estimate of the composite channel is given by
\begin{align}
		\widehat{\mathbf h}_u
		={}&
		\sqrt{\beta_u}\,\overline{\mathbf h}_u
		+\sqrt{q}\beta_u\mathbf R_u
		\left(q\beta_u\mathbf R_u+\sigma^2\mathbf I_M\right)^{-1}
		\nonumber\\
		&\times
		\left(
		\mathbf y_{p,u}
		-\sqrt{q\beta_u}\,\overline{\mathbf h}_u
		\right).
		\label{eq:general_mmse}
	\end{align}
The covariance matrices of the channel estimate and estimation error are, respectively, written as
\begin{align}
		\mathbf C_{\widehat h,u}
		&=
		q\beta_u^2\mathbf R_u
		\left(q\beta_u\mathbf R_u+\sigma^2\mathbf I_M\right)^{-1}
		\mathbf R_u,
		\label{eq:general_est_cov}\\
		\mathbf C_{\widetilde h,u}
		&=
		\beta_u\mathbf R_u-\mathbf C_{\widehat h,u}.
		\label{eq:general_error_cov}
\end{align}
	
Under the rich-scattering special case, the problem reduces to estimating $\mathbf g_u$ from the sufficient statistic in \eqref{eq:chan}. By using their joint Gaussian structure, the MMSE estimate is}
%By exploiting the joint Gaussian structure of $\mathbf{g}_u$ and $\mathbf{y}_{p,u}$, the linear MMSE estimator is obtained in closed form as
\begin{align}
	\hat{\mathbf{g}}_u
	=
	\frac{\sqrt{\tau_p \rho_p \beta_u}}
	{\tau_p \rho_p \beta_u + \sigma^2}
	\mathbf{y}_{p,u}.
\end{align}
Therefore, the estimation error is defined as
\begin{equation}
	\tilde{\mathbf{g}}_u
	=
	\mathbf{g}_u
	-
	\hat{\mathbf{g}}_u,
\end{equation}
where $\hat{\mathbf{g}}_u$ and $\tilde{\mathbf{g}}_u$ are statistically independent under Gaussian assumptions and their covariance matrices are given by
\begin{align}
	\mathbb{E}\{\hat{\mathbf{g}}_u \hat{\mathbf{g}}_u^{H}\}=
	\frac{\tau_p \rho_p \beta_u}
	{\tau_p \rho_p \beta_u + \sigma^2}
	\mathbf{I}_M
	\end{align}
	and
	\begin{align}
	\mathbb{E}\{\tilde{\mathbf{g}}_u \tilde{\mathbf{g}}_u^{H}\}
	=
	\frac{\sigma^2}
	{\tau_p \rho_p \beta_u + \sigma^2}
	\mathbf{I}_M.
\end{align}
Thus, the equivalent channel estimate is reconstructed as
\begin{align}
	\hat{\mathbf{h}}_u
	=
	\sqrt{\beta_u}\,
	\hat{\mathbf{g}}_u,
\end{align}
which yields the orthogonal decomposition
\begin{equation}
	\mathbf{h}_u
	=
	\hat{\mathbf{h}}_u
	+
	\tilde{\mathbf{h}}_u,
\end{equation}
where $\tilde{\mathbf{h}}_u = \sqrt{\beta_u}\tilde{\mathbf{g}}_u$. Consequently, the scalar estimation variances are given by
\begin{align}
	\Omega_{\hat{h},u}
	=
	\frac{\tau_p \rho_p \beta_u^2}
	{\tau_p \rho_p \beta_u + \sigma^2}\label{eq:qui}
\end{align}
and
\begin{align}
	\Omega_{\tilde{h},u}
	=
	\beta_u - \Omega_{\hat{h},u}.\label{eq:qui2}
\end{align}

These expressions reveal that the estimation accuracy improves with increasing pilot energy $\tau_p \rho_p$ and routing gain $\beta_u$. Specifically, stronger routing enhances the pilot observation quality and reduces the estimation error variance, improving the reliability of subsequent precoding operations.

\vspace{-2mm}
\subsection{Slow-Timescale Estimation of the Routing Gain}
The routing gain $\beta_u$ varies on a much slower timescale $T_r \gg T_c$, since it is determined by the large-scale SW routing geometry rather than the small-scale fading. Therefore, $\beta_u$ can be estimated using long-term averaging of the received pilot energy across multiple coherence blocks.

Let $B$ denote the number of coherence blocks used for averaging, and let $\mathbf{y}_{p,u}^{(b)}$ denote the pilot observation in block $b$. From the pilot model, we have
\begin{equation}
	\mathbf{y}_{p,u}^{(b)}
	=
	\sqrt{\tau_p \rho_p \beta_u}\,\mathbf{g}_u^{(b)}
	+
	\mathbf{n}_{p,u}^{(b)}.
\end{equation}
Under independent and identically distributed (i.i.d.)~Rayleigh fading, i.e., $\mathbf{g}_u \sim \mathcal{CN}(\mathbf{0},\mathbf{I}_M)$, we have
\begin{equation}
\mathbb{E}\{\|\mathbf{g}_u\|^2\} = M.
\end{equation}
Hence, the average received pilot energy satisfies
\begin{align}
	\mathbb{E}\{\|\mathbf{y}_{p,u}\|^2\}
	=
	\tau_p \rho_p \beta_u M
	+
	M \sigma^2.
\end{align}

An unbiased estimator of $\beta_u$ can thus be constructed as
\begin{align}\label{eq:beta_estimator}
\hat{\beta}_u=\frac{1}{B \tau_p \rho_p M}\sum_{b=1}^{B}\|\mathbf{y}_{p,u}^{(b)}\|^2-\frac{\sigma^2}{\tau_p \rho_p}. 
\end{align}

By the law of large numbers, $\hat{\beta}_u$ converges almost surely to $\beta_u$ as $B \rightarrow \infty$. This two-timescale separation enables infrequent estimation of the routing gain, while short-term fading estimation is performed per coherence block, thereby reducing training overhead and improving scalability. {\color{blue}Importantly, \eqref{eq:beta_estimator} reuses the pilot observations already acquired for short-term channel estimation and thus requires no additional pilot symbols beyond the per-block overhead $\tau_p/T_c$. Its cost is instead an acquisition latency. Let $T_{\rm cb}$ denote the duration of one coherence block. When the estimate is formed from the preceding $B$ blocks, the initial routing gain estimate becomes available after the duration
\begin{align}
T_{\beta}=BT_{\rm cb}.
\label{eq:routing_estimation_latency}
\end{align}
Thereafter, a sliding window can update the estimate once per coherence block without additional training transmissions.}

The above framework reveals a distinctive property of E-FAS-assisted systems: channel estimation quality depends not only on pilot resources $(\tau_p, \rho_p)$ but also on the routing gain $\beta_u$. Stronger SW routing increases the effective pilot observation SNR, thereby improving CSI accuracy in addition to enhancing the received signal power. Since $\beta_u$ evolves on a slow timescale, the proposed two-timescale structure reduces the need for frequent large-scale parameter learning.

\vspace{-2mm}
{\color{blue} \subsection{Finite Window Routing Gain Estimation and Mismatched MMSE Performance}
\label{subsec:finite_B_mismatch}

The almost sure convergence of the routing gain estimator in \eqref{eq:beta_estimator} holds as $B\rightarrow\infty$. For a finite averaging window, however, the estimate $\hat{\beta}_u$ remains random and may introduce a mismatch in the subsequent short-term channel estimator. Here, we characterize this effect exactly.

To begin with, define the accumulated pilot energy as
\begin{align}
q\triangleq\tau_p\rho_p.\label{eq:q_pilot_energy}
\end{align}
For the generalized channel model in \eqref{eq:general_channel}, conditioned on the normalized slow timescale spatial statistics $\overline{\mathbf h}_u$ and $\mathbf R_u$, the pilot observation satisfies
\begin{equation}
	\mathbf y_{p,u}^{(b)}
	\sim
	\mathcal{CN}\!\left(
	\sqrt{q\beta_u}\,\overline{\mathbf h}_u,
	\mathbf C_{p,u}
	\right),~\mbox{with }
	\mathbf C_{p,u}
	\triangleq
	q\beta_u\mathbf R_u+\sigma^2\mathbf I_M.
	\label{eq:general_pilot_distribution}
\end{equation}
Under the normalization $\|\overline{\mathbf h}_u\|^2+\operatorname{tr}(\mathbf R_u)=M$, the estimator in \eqref{eq:beta_estimator} remains unbiased, and its variance is
\begin{equation}
	\operatorname{Var}(\hat{\beta}_u)
	=
	\frac{
		\operatorname{tr}(\mathbf C_{p,u}^{2})
		+
		2q\beta_u
		\overline{\mathbf h}_u^{H}
		\mathbf C_{p,u}
		\overline{\mathbf h}_u
	}{
		Bq^2M^2
	}.
	\label{eq:beta_variance_general}
\end{equation}

%To obtain an exact finite sample distribution and a tractable mismatch characterization, the remainder of this subsection specializes to the rich scattering case $\overline{\mathbf h}_u=\mathbf 0$ and $\mathbf R_u=\mathbf I_M$.

Under rich scattering, each pilot observation satisfies 
\begin{align}
	\mathbf{y}_{p,u}^{(b)}
	\sim
	\mathcal{CN}
	\left(
	\mathbf{0},
	(q\beta_u+\sigma^2)\mathbf{I}_M
	\right).
	\label{eq:pilot_observation_distribution}
\end{align}
As such,
\begin{align}
	\frac{
		\sum_{b=1}^{B}
		\left\|
		\mathbf{y}_{p,u}^{(b)}
		\right\|^2
	}{
		q\beta_u+\sigma^2
	}
	\sim
	\Gamma(BM,1),
	\label{eq:pilot_energy_gamma}
\end{align}
which is parameterized by its shape and scale parameters.

Using \eqref{eq:beta_estimator}, the exact finite sample distribution
of the routing gain estimator is therefore given by
\begin{align}
	\hat{\beta}_u
	+
	\frac{\sigma^2}{q}
	\sim
	\Gamma\!\left(
	BM,
	\frac{q\beta_u+\sigma^2}{BMq}
	\right).
	\label{eq:beta_estimator_distribution}
\end{align}
It follows directly that
\begin{align}
	\mathbb{E}\{\hat{\beta}_u\}
	=
	\beta_u
	\label{eq:beta_estimator_mean}
\end{align}
and
\begin{align}
	\operatorname{Var}(\hat{\beta}_u)
	=
	\frac{
		(q\beta_u+\sigma^2)^2
	}{
		BMq^2
	}.
	\label{eq:beta_estimator_variance}
\end{align}

Accordingly, the relative root MSE (RMSE) of the routing gain estimator is given by
\begin{align}
	\eta_{\beta,u}(B)
	&\triangleq
	\frac{
		\sqrt{
			\mathbb{E}\{
			|\hat{\beta}_u-\beta_u|^2
			\}
		}
	}{
		\beta_u
	}
=
	\frac{
		1+\varrho_{p,u}^{-1}
	}{
		\sqrt{BM}
	},
	\label{eq:beta_relative_rmse}
\end{align}
where
\begin{equation}
	\varrho_{p,u}
	\triangleq
	\frac{q\beta_u}{\sigma^2}
	=
	\frac{\tau_p\rho_p\beta_u}{\sigma^2}
	\label{eq:effective_pilot_snr}
\end{equation}
denotes the effective pilot SNR for user $u$.

We next quantify the effect of routing gain mismatch on the
short-term MMSE estimator. Let $b_u\geq 0$ denote the routing-gain
value employed by the BS. The corresponding plug-in estimator of
the equivalent channel is
\begin{equation}
	\hat{\mathbf{h}}_u(b_u)
	=
	\frac{
		\sqrt{q}\,b_u
	}{
		qb_u+\sigma^2
	}
	\mathbf{y}_{p,u}.
	\label{eq:mismatched_channel_estimator}
\end{equation}
The resulting per-antenna estimation MSE is
\begin{align}
	\varepsilon_u(b_u)
	&\triangleq
	\frac{1}{M}
	\mathbb{E}
	\left\{
	\left\|
	\mathbf{h}_u
	-
	\hat{\mathbf{h}}_u(b_u)
	\right\|^2
	\right\}
	\nonumber\\
	&=
	\beta_u
	-
	\frac{
		2q\beta_u b_u
	}{
		qb_u+\sigma^2
	}
	+
	\frac{
		qb_u^2(q\beta_u+\sigma^2)
	}{
		(qb_u+\sigma^2)^2
	}.
	\label{eq:mismatched_mse_initial}
\end{align}
After straightforward algebraic manipulation,
\eqref{eq:mismatched_mse_initial} can equivalently be expressed as
\begin{align}
	\varepsilon_u(b_u)
	=
	\underbrace{
		\frac{
			\beta_u\sigma^2
		}{
			q\beta_u+\sigma^2
		}
	}_{\varepsilon_u^{\mathrm{MMSE}}}
	+
	\underbrace{
		\frac{
			q\sigma^4(b_u-\beta_u)^2
		}{
			(qb_u+\sigma^2)^2
			(q\beta_u+\sigma^2)
		}
	}_{\Delta\varepsilon_u(b_u)}.
	\label{eq:mismatched_mse_decomposition}
\end{align}
The first term in \eqref{eq:mismatched_mse_decomposition} is the matched MMSE value, whereas the second term is the non-negative
excess MSE caused by routing-gain mismatch. The mismatch penalty vanishes if and only if $b_u=\beta_u$.

When the finite window estimate is employed, we set $b_u=[\hat{\beta}_u]^+$, where $[x]^+\triangleq\max\{x,0\}$ ensures a physically valid nonnegative routing gain value.  Also, for normalized precoder construction, if $[\hat{\beta}_u]^+=0$, then an arbitrarily small positive floor $\epsilon_\beta>0$ can be used, i.e., $b_u=\max\{\hat{\beta}_u,\epsilon_\beta\}$. This avoids an undefined zero-norm channel estimate without changing the precoding direction, since any $b_u>0$ produces an estimate collinear with $\mathbf y_{p,u}$. The average
relative excess MSE is consequently defined as
\begin{align}
	\overline{\mathcal{L}}_{\mathrm{MSE},u}(B)
	\triangleq
	\mathbb{E}_{\hat{\beta}_u}
	\left\{
	\frac{
		\Delta\varepsilon_u([\hat{\beta}_u]^+)
	}{
		\varepsilon_u^{\mathrm{MMSE}}
	}
	\right\},
	\label{eq:average_mismatch_penalty}
\end{align}
where the expectation is evaluated with respect to the exact shifted Gamma distribution in \eqref{eq:beta_estimator_distribution}. Thus,
\eqref{eq:average_mismatch_penalty} can be evaluated through a $1$-D numerical integration without invoking an additional error distribution approximation.

For sufficiently large $BM$, a first-order expansion around $\hat{\beta}_u=\beta_u$ gives the useful scaling law
\begin{align}
	\overline{\mathcal{L}}_{\mathrm{MSE},u}(B)
	=
	\frac{1}{
		BM\varrho_{p,u}
	}
	+
	o\!\left(\frac{1}{B}\right).
	\label{eq:mismatch_penalty_scaling}
\end{align}
Consequently, the sensitivity to routing gain mismatch decreases with the number of BS antennas, the averaging window length, and the
effective pilot SNR.

\begin{remark}[Impact on Normalized MRT and ZF]
Although routing gain mismatch increases the channel estimation MSE, it does not affect the normalized MRT and ZF precoding directions under the i.i.d.~model considered in the closed-form performance analysis. To see this, we define
\begin{align}
a_u(b_u)\triangleq\frac{\sqrt{q}\,b_u}{qb_u+\sigma^2}.
\end{align}
For any $b_u>0$, the plug-in and matched MMSE estimates satisfy
\begin{align}
		\widehat{\mathbf h}_u(b_u)
		=
		\kappa_u(b_u)\widehat{\mathbf h}_u(\beta_u),~\mbox{with }
		\kappa_u(b_u)
		\triangleq
		\frac{a_u(b_u)}{a_u(\beta_u)}>0.
		\label{eq:estimate_scalar_relation}
	\end{align}
	Hence, the normalized MRT vector satisfies
	\begin{equation}
		\frac{\widehat{\mathbf h}_u(b_u)}
		{\|\widehat{\mathbf h}_u(b_u)\|}
		=
		\frac{\widehat{\mathbf h}_u(\beta_u)}
		{\|\widehat{\mathbf h}_u(\beta_u)\|}.
		\label{eq:mrt_mismatch_invariance}
	\end{equation}
	
	For the multiuser case, let
	$\mathbf K=\operatorname{diag}
	\{\kappa_1,\ldots,\kappa_K\}$.
	The mismatched and matched channel estimate matrices are related by
$
		\widehat{\mathbf H}(\mathbf b)
		=
		\widehat{\mathbf H}(\boldsymbol{\beta})\mathbf K
$. 
	Therefore, their unnormalized ZF beamforming matrices satisfy
	\begin{align}
		\mathbf V(\mathbf b)
		&=
		\widehat{\mathbf H}(\mathbf b)
		\left(
		\widehat{\mathbf H}^{H}(\mathbf b)
		\widehat{\mathbf H}(\mathbf b)
		\right)^{-1}=
		\mathbf V(\boldsymbol{\beta})\mathbf K^{-1},
		\label{eq:zf_mismatch_scaling}
	\end{align}
where the factor $\mathbf K^{-1}$ disappears when each ZF column is normalized to unit norm. As a consequence, under normalized MRT or ZF with fixed power allocation, we have
\begin{equation}
		P_{\mathrm{out}}^{(B),\mathrm{UB}}
		=
		P_{\mathrm{out}}^{\mathrm{UB}},
		\,\,\,
		R_{\mathrm{SU}}^{(B),\mathrm{LB}}
		=
		R_{\mathrm{SU}}^{\mathrm{LB}},
		\,\,\,
		R_{\mathrm{sum}}^{(B),\mathrm{LB}}
		=
		R_{\mathrm{sum}}^{\mathrm{LB}}.	\label{eq:rate_mismatch_invariance}
	\end{equation}
	%Thus, the finite window uncertainty quantified above affects channel reconstruction accuracy but not the considered normalized precoding performance. This invariance is specific to the uncorrelated scalar covariance model and does not generally hold 	for the correlated model in \eqref{eq:general_channel}.
Thus, under the considered fixed power and receiver-side CSI assumptions, finite window routing gain uncertainty affects the channel reconstruction MSE but not the post-precoding signal-to-interference plus noise ratio (SINR). Hence, outage, achievable rate, and bit error rate (BER) for any fixed modulation remain unchanged under the i.i.d.~model. This invariance does not generally hold for the correlated model in \eqref{eq:general_channel}.
\end{remark}

For a prescribed maximum relative routing gain RMSE $\eta_{\max}$, the minimum averaging window length is found by
\begin{align}
	B_u^\star
	\triangleq
	\min_{B\in\mathbb{N}}
	\left\{
	B:
	\eta_{\beta,u}(B)
	\leq
	\eta_{\max}
	\right\}.
	\label{eq:B_selection_problem}
\end{align}
For the generalized channel model, substituting \eqref{eq:beta_variance_general} into \eqref{eq:B_selection_problem} gives
\begin{align}
	B_{u,\mathrm{gen}}^\star
	=
	\left\lceil
	\frac{
		\operatorname{tr}(\mathbf C_{p,u}^{2})
		+
		2q\beta_u
		\overline{\mathbf h}_u^{H}
		\mathbf C_{p,u}
		\overline{\mathbf h}_u
	}{
		q^2\beta_u^2M^2\eta_{\max}^2
	}
	\right\rceil,
	\label{eq:B_selection_general}
\end{align}
%Using \eqref{eq:beta_relative_rmse}, the solution is available in closed-form as
such that under rich scattering, this reduces to
\begin{align}
	B_u^\star
	=
	\left\lceil
	\frac{
		(1+\varrho_{p,u}^{-1})^2
	}{
		M\eta_{\max}^2
	}
	\right\rceil.
	\label{eq:B_selection_closed_form}
\end{align}
%Since the routing gain is assumed approximately constant only over
%the routing timescale $T_r$, the selected window must additionally
%satisfy
%\begin{align}
%	B_u^\star T_c
%	\leq
%	T_r.
%	\label{eq:B_tracking_constraint}
%\end{align}
%If \eqref{eq:B_tracking_constraint} is violated, the requested
%accuracy cannot be attained solely by increasing $B$, since the
%resulting averaging interval would extend beyond the stationarity
%duration of the routing configuration.

%We define $T_{\rm cb}$ as the duration of one coherence block.
Since the routing gain is approximately constant only over the routing stationarity interval $T_r$, the selected averaging window must satisfy
\begin{align}
	B_u^\star
	\leq
	B_r
	\triangleq
	\left\lfloor
	\frac{T_r}{T_{\rm cb}}
	\right\rfloor.
	\label{eq:B_tracking_constraint}
\end{align}
If \eqref{eq:B_tracking_constraint} is violated, the prescribed accuracy cannot be achieved solely by increasing $B$, because the required averaging window would extend beyond the interval over which the routing gain remains approximately stationary.

The main analysis assumes that data transmission continues while the routing gain estimate is updated, and therefore, no additional pre-log penalty is introduced. For completeness, under a conservative cold-start protocol that suspends data transmission during the first $B$ blocks after each routing reconfiguration, the effective data fraction becomes
\begin{align}
	\zeta_{\rm cold}(B)
	=
	\left(1-\frac{B}{B_r}\right)^{+}
	\left(1-\frac{\tau_p}{T_c}\right).
	\label{eq:cold_start_prelog}
\end{align}
This factor represents a protocol upper bound on the latency penalty and is not required when an earlier estimate or online sliding window update is available.

Alternatively, when the design objective is stated directly in terms of the mismatched MMSE penalty, the minimum feasible window
can be selected as
\begin{align}
	B_{u,\mathrm{MSE}}^\star
	=
	\min_{\substack{
			B\in\mathbb{N}\\
		B\leq B_r
	}}
	\left\{
	B:
	\overline{\mathcal{L}}_{\mathrm{MSE},u}(B)
	\leq
	\mu_{\max}
	\right\},
	\label{eq:B_selection_mse}
\end{align}
in which $\mu_{\max}$ denotes the maximum tolerable relative excess MSE. Since \eqref{eq:average_mismatch_penalty} is a scalar
expectation, \eqref{eq:B_selection_mse} requires only a $1$-D integer search.}

\vspace{-2mm}
\section{Single-User Performance under Imperfect CSI}\label{sys-single}
%Here, we analyze the performance of an E-FAS-assisted system in the single-user case. The analysis explicitly accounts for the two-timescale channel structure and the MMSE estimation framework developed in Section~\ref{sec-pilot}.
{\color{blue}In this section, the single-user case provides an interference-free benchmark that isolates estimation error-induced self-interference and its fixed training high-SNR ceiling, thereby enabling the additional residual inter-user interference in the multiuser case to be identified separately.}

\vspace{-2mm}
\subsection{Transmission Model}
Consider the single-user case and omit the user index for notational clarity. The channel admits the decomposition
\begin{align}
\mathbf{h} = \hat{\mathbf{h}} + \tilde{\mathbf{h}},\label{eq:decom}
\end{align}
where $\hat{\mathbf{h}}$ and $\tilde{\mathbf{h}}$ denote the MMSE estimate  and the corresponding estimation error, respectively, which are statistically independent. Besides, the BS employs maximum ratio transmission (MRT) based on the estimated channel such that% \cite{Marzetta2016fund}
$
\mathbf{w} = \frac{\hat{\mathbf{h}}}{\|\hat{\mathbf{h}}\|}
$.  Therefore, the transmitted signal is $\mathbf{x}=\sqrt{P}\mathbf{w}x$ in which $\mathbb{E}\{|x|^2\}=1$. Hence, the received signal is given by
\begin{align}
y = \sqrt{P}\,\mathbf{h}^{H}\mathbf{w}x + n,
\end{align}
where $n \sim \mathcal{CN}(0,\sigma^2)$.

\vspace{-2mm}
%\subsection{Instantaneous SINR}
%Substituting the channel decomposition of \eqref{eq:decom} yields
%\begin{align}
%	y = \sqrt{P}\,\hat{\mathbf{h}}^{H}\mathbf{w}x
%	+ \sqrt{P}\,\tilde{\mathbf{h}}^{H}\mathbf{w}x
%	+ n.
%\end{align}
%Treating the unknown error term as additional Gaussian noise and conditioning on the channel estimate, the instantaneous SINR can be expressed as
%\begin{align}
%	\gamma =
%	\frac{P|\hat{\mathbf{h}}^{H}\mathbf{w}|^2}
%	{P\mathbb{E}\{|\tilde{\mathbf{h}}^{H}\mathbf{w}|^2 \mid \hat{\mathbf{h}}\} + \sigma^2}.
%\end{align}
%
%Since $\mathbf{w}$ is a deterministic function of $\hat{\mathbf{h}}$ and $\tilde{\mathbf{h}}$ is independent of $\hat{\mathbf{h}}$, the interference term becomes
%\begin{align}
%\mathbb{E}\{|\tilde{\mathbf{h}}^{H}\mathbf{w}|^2 \mid \hat{\mathbf{h}}\}
%=\mathbf{w}^{H}\mathbf{C}_{\tilde{h}}\mathbf{w},
%\end{align}
%where $\mathbf{C}_{\tilde{h}} = \mathbb{E}\{\tilde{\mathbf{h}}\tilde{\mathbf{h}}^{H}\}$ denotes the error covariance matrix.
%
%Under the uncorrelated Rayleigh fading assumption, i.e., $\mathbf{C}_{\tilde{h}} = \Omega_{\tilde{h}} \mathbf{I}_M$, and we therefore have
%\begin{align}
%\mathbb{E}\{|\tilde{\mathbf{h}}^{H}\mathbf{w}|^2\}=\Omega_{\tilde{h}}.
%\end{align}
%Moreover, MRT yields
%\begin{equation}
%|\hat{\mathbf{h}}^{H}\mathbf{w}|^2 = \|\hat{\mathbf{h}}\|^2.
%\end{equation}
%Hence, the instantaneous SINR simplifies to
%\begin{align}\label{eq:sinr}
%\gamma =\frac{P \|\hat{\mathbf{h}}\|^2}{P \Omega_{\tilde{h}} + \sigma^2}.
%\end{align}

{\color{blue}
\subsection{Conditional Effective SINR Lower Bound}

Substituting the channel decomposition in \eqref{eq:decom} into the received signal yields
\begin{align}
	y
	=
	\sqrt{P}\,\hat{\mathbf h}^{H}\mathbf w x
	+
	\sqrt{P}\,\tilde{\mathbf h}^{H}\mathbf w x
	+
	n.
\end{align}
It is assumed that the receiver knows the predicted effective channel coefficient $\hat{\mathbf h}^{H}\mathbf w$, but not the realization
of $\tilde{\mathbf h}^{H}\mathbf w$. Now, define the effective disturbance as
$
	v
	\triangleq
	\sqrt{P}\,\tilde{\mathbf h}^{H}\mathbf w x+n
$. 
Conditioned on $\hat{\mathbf h}$, this disturbance is zero mean and uncorrelated with the transmitted symbol, since
$
	\mathbb{E}
	\{
	x^{*}v\mid\hat{\mathbf h}
	\}
	=0$. 
Its conditional variance is given by
\begin{align}
	\mathbb{E}
	\left\{
	|v|^2\mid\hat{\mathbf h}
	\right\}
	=
	P\mathbf w^{H}\mathbf C_{\tilde h}\mathbf w+\sigma^2,
\end{align}
where $\mathbf C_{\tilde h}=\mathbb{E}\{\tilde{\mathbf h}\tilde{\mathbf h}^{H}\}$ is the channel estimation error covariance matrix.

Although $v$ is not generally independent Gaussian noise, the worst-case uncorrelated noise argument yields the conditional achievable rate lower bound
$
	I(x;y\mid\hat{\mathbf h})
	\geq
	\log_2\!\left(1+\gamma_{\rm LB}\right)$, 
where the corresponding effective SINR is
\begin{align}
	\gamma_{\rm LB}
	=
	\frac{
		P|\hat{\mathbf h}^{H}\mathbf w|^2
	}{
		P\mathbf w^{H}\mathbf C_{\tilde h}\mathbf w+\sigma^2
	}.
\end{align}

Under the uncorrelated Rayleigh fading assumption, $\mathbf C_{\tilde h}=\Omega_{\tilde h}\mathbf I_M$. Moreover, normalized MRT gives $|\hat{\mathbf h}^{H}\mathbf w|^2=\|\hat{\mathbf h}\|^2$. Therefore, we obtain
\begin{align}
	\gamma_{\rm LB}
	=
	\frac{
		P\|\hat{\mathbf h}\|^2
	}{
		P\Omega_{\tilde h}+\sigma^2
	},
	\label{eq:sinr}
\end{align}
which is the effective SINR associated with an achievable rate lower bound rather than the exact downlink SINR.}

\vspace{-2mm}
\subsection{Distribution of the Effective Channel Gain}
Under the two-timescale model, the estimated equivalent channel is expressed as
$
\hat{\mathbf{h}} = \sqrt{\beta}\,\hat{\mathbf{g}}
$. 
As a result, the MMSE estimate satisfies $
\hat{\mathbf{h}} \sim \mathcal{CN}(\mathbf{0}, \Omega_{\hat{h}}\mathbf{I}_M)$, 
where
$
\Omega_{\hat{h}}=\frac{\tau_p \rho_p \beta^2}{\tau_p \rho_p \beta + \sigma^2}$. 
Therefore, the entries of $\hat{\mathbf{h}}$ are i.i.d.~complex Gaussian random variables with variance $\Omega_{\hat{h}}$. Consequently, the squared norm
\begin{align}
\|\hat{\mathbf{h}}\|^2=\sum_{m=1}^{M} |\hat{h}_m|^2
\end{align}
is the sum of $M$ independent exponential random variables with mean $\Omega_{\hat{h}}$. Hence, $\|\hat{\mathbf{h}}\|^2$ follows a Gamma distribution with shape parameter $M$ and scale parameter $\Omega_{\hat{h}}$ \cite{Tulino2004random}, i.e.,
\begin{align}
\|\hat{\mathbf{h}}\|^2 \sim \Gamma(M,\Omega_{\hat{h}}).
\end{align}

Now, by letting $Z \triangleq \|\hat{\mathbf{h}}\|^2$, the corresponding probability density function (PDF) is given by
\begin{equation}
f_{Z}(z)=\frac{z^{M-1}}{\Gamma(M)\Omega_{\hat{h}}^{M}}\exp\!\left(-\frac{z}{\Omega_{\hat{h}}}\right),~z \ge 0,
\end{equation}
where $\Gamma(\cdot)$ denotes the Gamma function.

\vspace{-2mm}
{\color{blue}\subsection{Outage Probability Upper Bound}
%For a target spectral efficiency $R_{\mathrm{th}}$ (in bps/Hz), an outage event is defined as
%\begin{equation}
%P_{\mathrm{out}}\triangleq\Pr \left(\left(1-\frac{\tau_p}{T_c}\right)\log_2(1+\gamma)<R_{\mathrm{th}}\right),
%\end{equation}
%where $(1-\tau_p/T_c)$ accounts for the pilot overhead within each coherence block. Equivalently, the outage condition can be expressed in terms of an SINR threshold $\gamma_{\mathrm{th}}=2^{\frac{R_{\mathrm{th}}}{1-\tau_p/T_c}} - 1$, such that $P_{\mathrm{out}} = \Pr(\gamma < \gamma_{\mathrm{th}})$. Using the SINR expression in \eqref{eq:sinr}, the outage probability can be rewritten as
For a target spectral efficiency, $R_{\mathrm{th}}$ (in bps/Hz), we define the outage probability associated with the conditional rate lower bound as
\begin{align}
	P_{\mathrm{out}}^{\mathrm{UB}}
	\triangleq
	\Pr\left(
	\left(1-\frac{\tau_p}{T_c}\right)
	\log_2\left(1+\gamma_{\mathrm{LB}}\right)
	<R_{\mathrm{th}}
	\right).
\end{align}
Equivalently, it can be rewritten as 
$
	P_{\mathrm{out}}^{\mathrm{UB}}
	=
	\Pr\left(
	\gamma_{\mathrm{LB}}<\gamma_{\mathrm{th}}
	\right)$, 
in which 
$
	\gamma_{\mathrm{th}}
	=
	2^{\frac{R_{\mathrm{th}}}{1-\tau_p/T_c}}-1$. 
Since $\log_2(1+\gamma_{\mathrm{LB}})$ is a conditional achievable rate lower bound, $P_{\mathrm{out}}^{\mathrm{UB}}$ conservatively
upper bounds the actual outage probability under the stated receiver-side CSI assumption. Using \eqref{eq:sinr}, it follows that}
\begin{align}
P_{\mathrm{out}}^{\mathrm{UB}}
	=
	\Pr\left(
	\|\hat{\mathbf{h}}\|^2
	<
	\frac{(P\Omega_{\tilde{h}}+\sigma^2)\gamma_{\mathrm{th}}}{P}
	\right).
\end{align}

Since $\|\hat{\mathbf{h}}\|^2 \sim \Gamma(M,\Omega_{\hat{h}})$, the outage probability corresponds to the cumulative distribution function (CDF) of a Gamma random variable evaluated at $x_{\mathrm{th}}\triangleq\frac{(P\Omega_{\tilde{h}}+\sigma^2)\gamma_{\mathrm{th}}}{P}$. Therefore, it has the closed-form expression given by
\begin{align}
P_{\mathrm{out}}^{\mathrm{UB}}
	=
	\frac{
		\Upsilon\!\left(
		M,
		\frac{(P\Omega_{\tilde{h}}+\sigma^2)\gamma_{\mathrm{th}}}
		{P\Omega_{\hat{h}}}
		\right)
	}
	{\Gamma(M)},
\end{align}
where $\Upsilon(\cdot,\cdot)$ denotes the lower incomplete Gamma function.

\vspace{-2mm}
%\subsection{Achievable Rate with Training Overhead}
%Due to pilot transmission, only a fraction of each coherence block is available for data transmission. Accounting for this training overhead, the achievable ergodic rate is given by
%\begin{align}\label{eq:rate}
%R =\left(1-\frac{\tau_p}{T_c}\right)\mathbb{E}\{\log_2(1+\gamma)\},
%\end{align}
%where $\tau_p$ is the pilot length. Substituting the SINR expression from \eqref{eq:sinr} into \eqref{eq:rate}, the ergodic rate becomes
%\begin{equation}
%	R =
%	\left(1-\frac{\tau_p}{T_c}\right)
%	\mathbb{E}
%	\left[
%	\log_2
%	\left(
%	1+
%	\frac{P \|\hat{\mathbf{h}}\|^2}
%	{P\Omega_{\tilde{h}}+\sigma^2}
%	\right)
%	\right].
%\end{equation}
%Since $\|\hat{\mathbf{h}}\|^2 \sim \Gamma(M,\Omega_{\hat{h}})$, the expectation can be found as
%\begin{align}\notag
%	R = &
%	\frac{1-\frac{\tau_p}{T_c}}{\ln 2}\\
%	&\times
%	\int_{0}^{\infty}
%	\ln\!\left(
%	1+
%	\frac{P x}
%	{P\Omega_{\tilde{h}}+\sigma^2}
%	\right)
%	\frac{x^{M-1}}
%	{\Gamma(M)\Omega_{\hat{h}}^{M}}
%	\exp\!\left(-\frac{x}{\Omega_{\hat{h}}}\right)
%	dx\label{eq:int}\\
%	&=\left(1-\frac{\tau_p}{T_c}\right)
%	\frac{e^{z}}{\ln 2}
%	\sum_{k=1}^{M} \mathrm{E}_k(z),
%	\label{eq:R_closed_form}
%\end{align}
%where $z=\frac{P\Omega_{\tilde{h}}+\sigma^2}{P\,\Omega_{\hat{h}}}$ and $\mathrm{E}_n(z)\triangleq \int_{1}^{\infty}\frac{e^{-zt}}{t^n}\,dt$ for $z>0$ is the generalized exponential integral of order $n$. The proof of \eqref{eq:R_closed_form} is given in Appendix~\ref{app:rate_integral}.

{\color{blue}
\subsection{Achievable Ergodic Rate Lower Bound}
Accounting for the pilot overhead, an achievable ergodic rate lower bound can be found as
\begin{align}
	R_{\rm SU}^{\rm LB}
	=
	\left(1-\frac{\tau_p}{T_c}\right)
	\mathbb{E}
	\left[
	\log_2\left(1+\gamma_{\rm LB}\right)
	\right].
	\label{eq:su_achievable_rate}
\end{align}
Substituting \eqref{eq:sinr} into
\eqref{eq:su_achievable_rate} gives
\begin{equation}
	R_{\rm SU}^{\rm LB}
	=
	\left(1-\frac{\tau_p}{T_c}\right)
	\mathbb{E}
	\left[
	\log_2
	\left(
	1+
	\frac{P\|\hat{\mathbf h}\|^2}
	{P\Omega_{\tilde h}+\sigma^2}
	\right)
	\right].
\end{equation}
Since $\|\hat{\mathbf h}\|^2\sim\Gamma(M,\Omega_{\hat h})$, the expectation can be written as
\begin{align}
	R_{\rm SU}^{\rm LB}
	={}&
	\frac{1-\frac{\tau_p}{T_c}}{\ln 2}
	\int_{0}^{\infty}
	\ln\left(
	1+
	\frac{Px}
	{P\Omega_{\tilde h}+\sigma^2}
	\right)
	\nonumber\\
	&\times
	\frac{x^{M-1}}
	{\Gamma(M)\Omega_{\hat h}^{M}}
	\exp\left(-\frac{x}{\Omega_{\hat h}}\right)
	dx
	\label{eq:int}\\
	={}&
	\left(1-\frac{\tau_p}{T_c}\right)
	\frac{e^{z}}{\ln 2}
	\sum_{k=1}^{M}\mathrm{E}_k(z),
	\label{eq:su_rate_closed_form}
\end{align}
where
$
	z
	=
	\frac{P\Omega_{\tilde h}+\sigma^2}
	{P\Omega_{\hat h}}$, 
and $\mathrm{E}_n(z)\triangleq\int_{1}^{\infty}e^{-zt}t^{-n}\,dt$ for $z>0$ is the generalized exponential integral of order $n$. The proof of \eqref{eq:su_rate_closed_form} is given in Appendix~\ref{app:rate_integral}.}

\vspace{-2mm}
\subsection{High-SNR Behavior}\label{sec:high2}
%As $P \rightarrow \infty$, the instantaneous SINR in \eqref{eq:sinr} converges to
%\begin{equation}
%	\gamma_{\infty}
%	=
%	\frac{\|\hat{\mathbf{h}}\|^2}
%	{\Omega_{\tilde{h}}}.
%\end{equation}
%Unlike the perfect CSI case, where $\gamma$ grows unboundedly with $P$, 
%the imperfect CSI system exhibits SINR saturation due to the residual self-interference term that scales linearly with $P$. Accordingly, the achievable rate converges to the finite limit
%\begin{equation}
%	R_{\infty}
%	=
%	\left(1-\frac{\tau_p}{T_c}\right)
%	\mathbb{E}
%	\left[
%	\log_2
%	\left(
%	1+
%	\frac{\|\hat{\mathbf{h}}\|^2}{\Omega_{\tilde{h}}}
%	\right)
%	\right].
%\end{equation}
%
%Importantly, the estimation variances $\Omega_{\hat{h}}$ and $\Omega_{\tilde{h}}$ depend explicitly on the routing gain $\beta$. A larger $\beta$ improves the effective pilot SNR, thereby reducing $\Omega_{\tilde{h}}$ and increasing the saturation level. Hence, SW-assisted propagation improves not only the received signal strength but also the robustness of CSI acquisition. This intrinsic coupling between propagation geometry and estimation accuracy highlights a distinctive feature of E-FAS-assisted systems, where the topology-dependent routing gain directly governs both received signal strength and CSI quality. In contrast to conventional space wave MIMO systems with fixed large-scale fading, the routing configuration in E-FAS influences the effective pilot observation SNR and thus the estimation-induced interference level.

{\color{blue}
Let $\rho_d\triangleq\frac{P}{\sigma^2}$, $\varrho_p\triangleq\frac{q\beta}{\sigma^2}$, and $X_M\triangleq\|\widehat{\mathbf h}\|^2/
\Omega_{\widehat h}\sim\Gamma(M,1)$. Hence, the single-user SINR can be written as
\begin{align}
	\gamma^\mathrm{LB}
	=
	\frac{
		\rho_d\beta\varrho_p
	}{
		\rho_d\beta+1+\varrho_p
	}
	X_M.
	\label{eq:su_snr_scaling}
\end{align}
For fixed pilot resources, $\varrho_p$ remains bounded as $\rho_d\rightarrow\infty$, and therefore, $\gamma^\mathrm{LB}\longrightarrow\varrho_p X_M$. Thus, the previously identified ceiling is specifically a fixed training high-SNR result. More generally, suppose that $\varrho_p=c\rho_d^{\alpha}$ for some $c>0$ and $\alpha\geq0$. Then we get
\begin{align}
	\gamma^\mathrm{LB}
	\sim
	\begin{cases}
		cX_M, & \alpha=0,\\
		c\rho_d^{\alpha}X_M, & 0<\alpha<1,\\
		\dfrac{\beta c}{\beta+c}\rho_d X_M, & \alpha=1,\\
		\beta\rho_d X_M, & \alpha>1.
	\end{cases}
	\label{eq:su_pilot_scaling}
\end{align}
Consequently,
\begin{align}
\lim_{\rho_d\rightarrow\infty}
\frac{R_{\mathrm{SU}}^{\mathrm{LB}}}
{\log_2\rho_d}
=
\left(1-\frac{\tau_p}{T_c}\right)\min\{\alpha,1\}.
	\label{eq:su_rate_slope}
	\end{align}
As such, the routing gain $\beta$ improves both the received signal strength and the effective pilot quality, while a finite SINR ceiling arises only when the accumulated pilot SNR remains bounded.}

{\color{blue}We define the diversity order from the outage upper bound as
	$
	d_{\rm out}\triangleq
	-\lim_{\rho_d\rightarrow\infty}
	\frac{\log P_{\rm out}^{\rm UB}}{\log\rho_d}
$. Using the small argument behavior of the Gamma CDF with $\varrho_p=c\rho_d^\alpha$, we obtain
\begin{align}
d_{\rm out}=M\min\{\alpha,1\}.
\end{align}
Hence, fixed pilot resources $(\alpha=0)$ produce an outage floor, which has zero diversity order, whereas the full order $M$ is recovered when the effective pilot SNR scales at least linearly with the downlink SNR.}
\vspace{-2mm}
\section{Multiuser Transmission under Imperfect CSI}\label{sec:mult}
Consider the downlink transmission to $K$ single-antenna users served by an $M$-antenna BS, where $K < M$. The analysis explicitly accounts for the two-timescale structure and the MMSE estimation framework developed in Section~\ref{sec-pilot}. {\color{blue}Here, ZF is adopted because it suppresses inter-user interference with respect to the estimated channels, thereby isolating the residual interference caused by channel estimation errors.}

\vspace{-2mm}
\subsection{ZF Precoding Based on Channel Estimates}
Let $\hat{\mathbf{H}}=\big[ \hat{\mathbf{h}}_{1}, \ldots, \hat{\mathbf{h}}_{K} \big]\in \mathbb{C}^{M \times K}$ represent the estimated channel matrix. Provided that $M > K$, the BS constructs the ZF precoder as \cite{bjo2017massivem}
\begin{align}
\mathbf{W}=\hat{\mathbf{H}}\left(\hat{\mathbf{H}}^{H}\hat{\mathbf{H}}\right)^{-1}\mathbf{D},
\end{align}
where the diagonal matrix $\mathbf{D} = \mathrm{diag}\{d_1,\ldots,d_K\}$ is chosen such that $\|\mathbf{w}_u\|^2 = 1$ for all $u$, ensuring that
\begin{align}
\mathbb{E}\{\|\mathbf{x}\|^2\}=\sum_{u=1}^K P_u=P.
\end{align}

We also define $\mathbf{w}_u$ as the $u$-th column of $\mathbf{W}$. Hence, the transmitted signal is given by
\begin{align}
\mathbf{x}=\sum_{u=1}^{K}\sqrt{P_u}\mathbf{w}_u x_u,
\end{align}
where $\mathbb{E}\{|x_u|^2\}=1$. 

\vspace{-2mm}
\subsection{Received Signal Model}
The true channel admits the decomposition
\begin{equation}
\mathbf{h}_u = \hat{\mathbf{h}}_u + \tilde{\mathbf{h}}_u,
\end{equation}
where $\hat{\mathbf{h}}_u$ and $\tilde{\mathbf{h}}_u$ are statistically independent. The received signal at user $u$ is given by
\begin{align}
	y_u
	&=
	\mathbf{h}_u^{H}\mathbf{x} + n_u 
	\notag\\
	&=
	\sum_{i=1}^{K}
	\sqrt{P_i}
	\hat{\mathbf{h}}_u^{H}\mathbf{w}_i x_i
	+
	\sum_{i=1}^{K}
	\sqrt{P_i}
	\tilde{\mathbf{h}}_u^{H}\mathbf{w}_i x_i
	+
	n_u.
\end{align}
Since ZF precoding ensures that $\hat{\mathbf{h}}_u^{H}\mathbf{w}_i = 0$ for $i \neq u$, the received signal reduces to
\begin{multline}\label{eq-yum}
y_u=\sqrt{P_u}\hat{\mathbf{h}}_u^{H}\mathbf{w}_u x_u+\sqrt{P_u}\tilde{\mathbf{h}}_u^{H}\mathbf{w}_u x_u\\
+\sum_{i\neq u}\sqrt{P_i}\tilde{\mathbf{h}}_u^{H}\mathbf{w}_i x_i+n_u,
\end{multline}
where the first term represents the desired signal component formed along the estimated channel direction. The second term corresponds to residual self-interference due to estimation errors, while the third term represents multiuser interference caused by imperfect channel knowledge.

\vspace{-2mm}
%\subsection{SINR Expression}
%Due to channel estimation errors, the BS performs ZF precoding based on $\hat{\mathbf{H}}$ instead of the true channel $\mathbf{H}$. Using the standard use-and-forget bound \cite{bjo2017massivem}, we treat the unknown channel estimation errors as additional Gaussian noise and condition on the channel estimates. Accordingly, the effective SINR at user $u$ can be expressed as
%\begin{align}
%	\gamma_u
%	=
%	\frac{P_u |\hat{\mathbf{h}}_u^{H}\mathbf{w}_u|^2}
%	{
%		\sum_{i=1}^{K} P_i
%		\mathbb{E}\!\left\{
%		|\tilde{\mathbf{h}}_u^{H}\mathbf{w}_i|^2
%		\,\big|\,
%		\hat{\mathbf{H}}
%		\right\}
%		+
%		\sigma^2}.
%\end{align}

\subsection{Conditional Effective SINR Lower Bound}
{\color{blue}Due to channel estimation errors, the BS performs ZF precoding based on $\hat{\mathbf H}$ instead of the true channel
$\mathbf H$. Conditioned on $\hat{\mathbf H}$, we assume that user $u$ knows the predicted effective coefficient $\hat{\mathbf h}_u^H\mathbf w_u$, but not the instantaneous estimation error terms. Applying the conditional worst-case uncorrelated noise argument \cite{bjo2017massivem}, the effective SINR associated with an achievable rate lower bound is given by}
\begin{align}
	\gamma_u
	=
	\frac{P_u |\hat{\mathbf{h}}_u^{H}\mathbf{w}_u|^2}
	{
		\sum_{i=1}^{K} P_i
		\mathbb{E}\!\left\{
		|\tilde{\mathbf{h}}_u^{H}\mathbf{w}_i|^2
		\,\big|\,
		\hat{\mathbf{H}}
		\right\}
		+
		\sigma^2}.
\end{align}

Given that $\tilde{\mathbf{h}}_u$ is independent of $\hat{\mathbf{H}}$, and $\mathbf{w}_i$ is a deterministic function of $\hat{\mathbf{H}}$, the conditional interference term becomes
\begin{equation}
	\mathbb{E}\!\left\{
	|\tilde{\mathbf{h}}_u^{H}\mathbf{w}_i|^2
	\,\big|\,
	\hat{\mathbf{H}}
	\right\}
	=
	\mathbf{w}_i^{H}
	\mathbf{C}_{\tilde{h},u}
	\mathbf{w}_i,
\end{equation}
in which $\mathbf{C}_{\tilde{h},u} = \mathbb{E}\{\tilde{\mathbf{h}}_u \tilde{\mathbf{h}}_u^{H}\}$. Under the fading assumption $\mathbf{C}_{\tilde{h},u} = \Omega_{\tilde{h},u}\mathbf{I}_M$ and since $\|\mathbf{w}_i\|^2=1$, we obtain
\begin{align}
	\mathbb{E}\!\left\{
	|\tilde{\mathbf{h}}_u^{H}\mathbf{w}_i|^2
	\,\big|\,
	\hat{\mathbf{H}}
	\right\}
	=
	\Omega_{\tilde{h},u}.
\end{align}
Hence, the SINR becomes
\begin{align}
	\gamma_u
	=
	\frac{P_u |\hat{\mathbf{h}}_u^{H}\mathbf{w}_u|^2}
	{
		\Omega_{\tilde{h},u} \sum_{i=1}^{K} P_i
		+
		\sigma^2
	}.
\end{align}

It should be noted that in the E-FAS setting both the desired signal term and the interference components depend explicitly on the topology-dependent routing gain $\beta$, through the estimation variances $\Omega_{\hat{h}}$ and $\Omega_{\tilde{h}}$. This dependence fundamentally governs the high-SNR behavior, to be analyzed next.

\vspace{-2mm}
\subsection{Distribution of the Effective Channel Gain}
Under the two-timescale channel model,
$
\hat{\mathbf{h}}_u = \sqrt{\beta_u}\,\hat{\mathbf{g}}_u
$, 
and uncorrelated Rayleigh fading $\mathbf{R}_u=\mathbf{I}_M$, the channel estimate satisfies
$
\hat{\mathbf{h}}_u \sim\mathcal{CN}(\mathbf{0},\Omega_{\hat{h},u}\mathbf{I}_M)
$. 
Since the ZF precoder nulls the estimated channels of the other $K-1$ users, the effective beamforming vector $\mathbf{w}_u$ lies in an $(M-K+1)$-dimensional subspace orthogonal to $\{\hat{\mathbf{h}}_i\}_{i \neq u}$. Thus, the effective channel gain $|\hat{\mathbf{h}}_u^{H}\mathbf{w}_u|^2$ is Gamma distributed with shape parameter $M-K+1$ and scale parameter $\Omega_{\hat{h},u}$ \cite{Tulino2004random,wagner2012large}, i.e.,
\begin{align}
Z_u \triangleq |\hat{\mathbf{h}}_u^{H}\mathbf{w}_u|^2\sim \Gamma(M-K+1,\Omega_{\hat{h},u}).
\end{align}

Now, assuming equal power allocation $P_u=P/K$, the SINR simplifies to
\begin{align}
\gamma_u=\frac{\frac{P}{K} Z_u}{P\Omega_{\tilde{h},u}+\sigma^2},
\end{align}
where $Z_u \sim \Gamma(M-K+1,\Omega_{\hat{h},u})$.

\vspace{-2mm}
\subsection{Sum-Rate Lower Bound with Training Overhead}
Accounting for pilot overhead within each coherence block of length $T_c$, an ergodic sum-rate lower bound is found as
\begin{equation}
{\color{blue}R_{\mathrm{sum}}^\mathrm{LB}}=\left(1-\frac{\tau_p}{T_c}\right)\sum_{u=1}^{K}\mathbb{E}\{\log_2(1+\gamma_u)\}.
\end{equation}
Under statistical symmetry among users and equal power allocation $P_u=P/K$, all users experience identical SINR statistics. Hence, the sum-rate can be defined as
\begin{equation}
{\color{blue}R_{\mathrm{sum}}^\mathrm{LB}}\triangleq\left(1-\frac{\tau_p}{T_c}\right)K\mathbb{E}\left[\log_2\left(1+\frac{\frac{P}{K} Z}{P\Omega_{\tilde{h}}+\sigma^2}\right)\right],
\end{equation}
where $Z \sim \Gamma(M-K+1,\Omega_{\hat{h}})$.

Assuming $a \triangleq \frac{P/K}{P\Omega_{\tilde{h}}+\sigma^2}$, the expectation can be written as
\begin{align}\notag
	&{\color{blue}R_{\mathrm{sum}}^\mathrm{LB}}
	=
	\left(1-\frac{\tau_p}{T_c}\right)
	\frac{K}{\ln 2}
	\\
	&\times\int_{0}^{\infty}
	\ln(1+a x)
	\frac{x^{M-K}}
	{\Gamma(M-K+1)\Omega_{\hat{h}}^{M-K+1}}
	\exp\!\left(-\frac{x}{\Omega_{\hat{h}}}\right)
	dx.
\end{align}
Using standard integral identities for Gamma-distributed random variables, the closed-form expression is obtained as
\begin{align}
	{\color{blue}R_{\mathrm{sum}}^\mathrm{LB}}
	=
	\left(1-\frac{\tau_p}{T_c}\right)
	\frac{K}{\ln 2}
	\exp(z)
	\sum_{m=1}^{M-K+1}
	\mathrm{E}_m(z),\label{eq:r-multi}
\end{align}
in which $z=\frac{K(P\Omega_{\tilde{h}}+\sigma^2)}{P\Omega_{\hat{h}}}$. The proof follows from the same analytical steps as in the single-user case, with the Gamma shape parameter adjusted from $M$ to $M-K+1$ to account for the ZF-induced dimensionality reduction.

{\color{blue}
\vspace{-2mm}
\subsection{High-SNR Analysis}\label{sec:high}

We define
$
	\rho_d\triangleq\frac{P}{\sigma^2}$, $
	\varrho_{p,u}\triangleq\frac{q\beta_u}{\sigma^2}
$, 
and 
$
	X_{M-K+1}
	\triangleq
	\frac{Z_u}{\Omega_{\hat h,u}}
	\sim\Gamma(M-K+1,1)$. 
Therefore, under equal power allocation, the user SINR can be expressed as
\begin{align}
	\gamma_u
	=
	\frac{
		\rho_d\beta_u\varrho_{p,u}
	}{
		K\left(
		\rho_d\beta_u+1+\varrho_{p,u}
		\right)
	}
	X_{M-K+1}.
	\label{eq:mu_snr_scaling}
\end{align}

For fixed pilot resources, $\varrho_{p,u}$ remains bounded as
$\rho_d\rightarrow\infty$, and hence
$
	\gamma_u
	\longrightarrow
	\frac{\varrho_{p,u}}{K}X_{M-K+1}.
%	\label{eq:mu_fixed_training_ceiling}
$
Accordingly, under statistical symmetry, the sum-rate converges to
\begin{equation}
	R_{\mathrm{sum},\infty}^\mathrm{LB}
	=
	\left(1-\frac{\tau_p}{T_c}\right)
	K\,
	\mathbb{E}
	\left[
	\log_2
	\left(
	1+\frac{\varrho_p}{K}X_{M-K+1}
	\right)
	\right].
	\label{eq:mu_fixed_training_rate_ceiling}
\end{equation}
As a result, the finite ceiling is specifically a fixed training high-SNR result. More generally, suppose that $\varrho_{p,u}=c_u\rho_d^{\alpha}$, where $c_u>0$ and $\alpha\geq0$. Then we get
\begin{align}
	\gamma_u
	=
	\Theta\!\left(
	\rho_d^{\min\{\alpha,1\}}
	\right),
\end{align}
and therefore, 
\begin{align}
	\lim_{\rho_d\rightarrow\infty}
	\frac{{\color{blue}R_{\mathrm{sum}}^\mathrm{LB}}}{\log_2\rho_d}
	=
	\left(1-\frac{\tau_p}{T_c}\right)
	K\min\{\alpha,1\}.
	\label{eq:mu_rate_slope}
\end{align}
Consequently, pilot power scaling removes the ZF sum-rate ceiling for $\alpha>0$, while the full multiplexing slope is also recovered in the case of $\alpha\geq1$.}

{\color{blue}
\subsection{Joint Pilot Power and Pilot Length Allocation}\label{subsec:pilot_resource_allocation}
The channel estimation quality depends on the accumulated pilot energy $\tau_p\rho_p$, whereas the effective data transmission duration decreases with $\tau_p$. Therefore, selecting the pilot power and length independently may result in either insufficient channel estimation accuracy or excessive training overhead.

Now, we denote $\mathcal{R}_j(\tau_p,\rho_p)$ as the effective single-user rate for $j=\mathrm{SU}$ or the multiuser sum-rate for $j=\mathrm{MU}$, as given by \eqref{eq:su_rate_closed_form} and \eqref{eq:r-multi}, respectively. We consider a maximum pilot energy budget $E_{p,\max}$ per coherence block and a peak per-symbol pilot power constraint $\rho_{p,\max}$. The joint allocation problem is formulated as
\begin{align}
\max_{\tau_p,\rho_p}
	\quad&
	\mathcal{R}_j(\tau_p,\rho_p)
	\label{eq:joint_pilot_optimization}\\
	\mbox{s.t.}
	\quad&
	\tau_p\rho_p\leq E_{p,\max},
	\nonumber\\
	&
	0\leq\rho_p\leq\rho_{p,\max},
	\nonumber\\
	&
	\tau_p\in\mathcal{T}_j,
	\nonumber
\end{align}
where
$
	\mathcal{T}_{\mathrm{SU}}
	=
	\{1,\ldots,T_c-1\}
$
and
$
	\mathcal{T}_{\mathrm{MU}}
	=
	\{K,\ldots,T_c-1\}$. 
The lower bound $\tau_p\geq K$ in the multiuser case ensures the availability of mutually orthogonal pilot sequences.

For a fixed $\tau_p$, both $\Omega_{\widehat{h},u}$ and the resulting effective SINR increase monotonically with the accumulated pilot energy $\tau_p\rho_p$, whereas $\Omega_{\widetilde{h},u}$ decreases monotonically. Thus, the largest feasible pilot power is optimal for each candidate pilot length:
\begin{align}
	\rho_p^\star(\tau_p)
	=
	\min\left\{
	\rho_{p,\max},
	\frac{E_{p,\max}}{\tau_p}
	\right\}.
	\label{eq:optimal_power_given_length}
\end{align}
Substituting \eqref{eq:optimal_power_given_length} into
\eqref{eq:joint_pilot_optimization} reduces the joint
mixed integer problem to
\begin{align}
	\tau_{p,j}^\star
	=
	\underset{\tau_p\in\mathcal{T}_j}
	{\operatorname{arg\,max}}\;
	\mathcal{R}_j
	\left(
	\tau_p,
	\rho_p^\star(\tau_p)
	\right),
	\label{eq:optimal_pilot_length}
\end{align}
followed by
\begin{align}
	\rho_{p,j}^\star
	=
	\rho_p^\star(\tau_{p,j}^\star).
	\label{eq:optimal_pilot_power}
\end{align}
Since $\mathcal{T}_j$ is a finite integer set, \eqref{eq:optimal_pilot_length} provides the globally optimal allocation within the stated energy and peak power constraints through a $1$-D search. Since $E_{p,\max}$ and $\rho_{p,\max}$ are fixed independently of $\rho_d$, the optimized pilot energy remains bounded and corresponds to the $\alpha=0$ regime in Sections \ref{sec:high2} and \ref{sec:high}. A larger routing gain $\beta_u$ increases the effective pilot SNR for the same pilot resources. Consequently, the marginal benefit of extending the training interval generally decreases with $\beta_u$, which may shift the optimal pilot length toward a smaller value in E-FAS-assisted transmission.}

\vspace{-2mm}
\section{Numerical Results}\label{sec:num}
In this section, we evaluate the performance of E-FAS-assisted downlink transmission under imperfect CSI and validate the analytical expressions derived in the previous sections. The numerical results aim to quantify the impact of SW-assisted propagation, channel estimation quality, and pilot overhead on outage probability and achievable rate.

Unless otherwise specified, the simulation parameters are set as follows. Each Monte Carlo point is obtained by averaging over $2\times10^{4}$ independent channel realizations. The BS is equipped with $M=64$ antennas. For the multiuser case, the BS serves $K=8$ single-antenna users. The channel coherence block length is $T_c=200$ symbols and the noise variance is normalized to $\sigma^2=1$. Uplink training is performed using orthogonal pilots with per-symbol pilot SNR $\rho_p/\sigma^2=0$~dB, i.e., $\rho_p=1$. The pilot length is set to $\tau_p = K$ in the multiuser case, as commonly adopted for orthogonal pilot transmission in TDD multiuser MIMO systems \cite{ngo2013energy}. %while in the single-user case, $\tau_p = 8$ is adopted as a representative operating point that provides a reasonable trade-off between channel estimation accuracy and pilot overhead, unless stated otherwise.
 {\color{blue} For figures intended to isolate the effects of routing gain,
 	estimator type, or downlink SNR, $\tau_p=8$ is used only as a
 	common reference setting and is not intended to be optimal. The
 	jointly optimal pilot length and power are obtained separately from
 	the pilot resource allocation problem developed in Section \ref{subsec:pilot_resource_allocation}.
 	For the finite window routing gain analysis, the averaging window length is varied over $B\in\{1,\ldots,20\}$, and a target relative routing gain RMSE of $\eta_{\max}=5\%$ is considered.} The downlink SNR is defined as $\rho_d \triangleq P/\sigma^2$ and is varied over $[-10,30]$~dB. Furthermore, the equivalent end-to-end channel is modeled as $\mathbf{h}_{u}\sim\mathcal{CN}(\mathbf{0},\beta\mathbf{I}_M)$, in which $\beta$ represents the large-scale channel power capturing the end-to-end propagation gain. {\color{blue}The reported routing gains are normalized by the large-scale gain of the reference conventional link; thus, $\beta_u=1$ denotes the baseline case, whereas $\beta_u>1$ represents a relative E-FAS routing enhancement rather than an absolute passive power gain.} For simplicity and to isolate the impact of CSI quality, symmetric large-scale conditions are assumed for all users.  Unless otherwise stated, we assume $\beta_u=\beta$ for all users.  Additionally, to comprehensively evaluate the influence of CSI accuracy on system performance, we consider the following benchmark schemes:  (i) Ideal transmission assuming perfect CSI; (ii) Imperfect CSI based on MMSE channel estimation; (iii) Imperfect CSI based on LS channel estimation; and (iv) Conventional transmission scheme without E-FAS.

Additionally, to ensure a physically meaningful comparison between the E-FAS-assisted and conventional transmission scenarios, we consider a representative large-scale propagation geometry. Specifically, a direct BS-user link of distance $d_0$ is assumed for the conventional case, whereas the E-FAS-assisted route consists of two segments with distances $d_1$ (BS-E-FAS) and $d_2$ (E-FAS-user). The large-scale channel gain is modeled using a standard distance-dependent path-loss model $\beta(d) = C_0 \left( \frac{d}{d_0} \right)^{-\alpha}$, where $\alpha$ denotes the path-loss exponent and $C_0$ is chosen such that the direct link satisfies $\beta_{\text{no}}=1$ at distance $d_0$. The composite gain of the E-FAS-assisted path therefore depends on the relative segment distances and $\alpha$, and can yield a moderate large-scale enhancement (on the order of a few dB) under representative configurations. In the subsequent analysis, $\beta$ is treated parametrically for analytical tractability, while the above geometry provides a physically grounded interpretation of the adopted gain values.
{\color{blue}Moreover, to quantify the performance overestimation caused by neglecting channel estimation errors, we define the relative perfect CSI overestimation for a performance metric $X$ as
$
	\mathcal{L}_{X}
	\triangleq
	\frac{
		X_{\mathrm{perfect}}
		-
		X_{\mathrm{MMSE}}
	}{
		X_{\mathrm{perfect}}
	}
	\times 100\%,
	%\label{eq:relative_csi_loss}
$
where $X$ denotes either the single-user achievable rate bound or the multiuser ergodic sum-rate. This metric directly quantifies the error incurred when conclusions obtained under perfect CSI are applied to a system employing practical pilot channel estimation.}

\begin{table}[t]
	\centering
	{\color{blue}\caption{Relative performance overestimation under perfect CSI for the E-FAS setting with $M=64$, $\beta=3$, $T_c=200$, $\tau_p=8$, $\rho_p/\sigma^2=0$ dB, and $K=8$ in the multiuser case.}
	\label{tab:csi_overestimation}
	\begin{tabular}{c||c|c}
		\hline
		Downlink SNR
		&
		SU rate
		&
		MU sum-rate
		\\
		(dB) & overestimation & overestimation
		\\
		\hline
		$10$ & $11.0\%$ & $15.4\%$ \\
		$20$ & $26.4\%$ & $33.9\%$ \\
		$30$ & $39.8\%$ & $48.5\%$ \\
		\hline
	\end{tabular}}\vspace{-5mm}
\end{table}

\begin{figure}[]
\centering
\includegraphics[width=0.9\columnwidth]{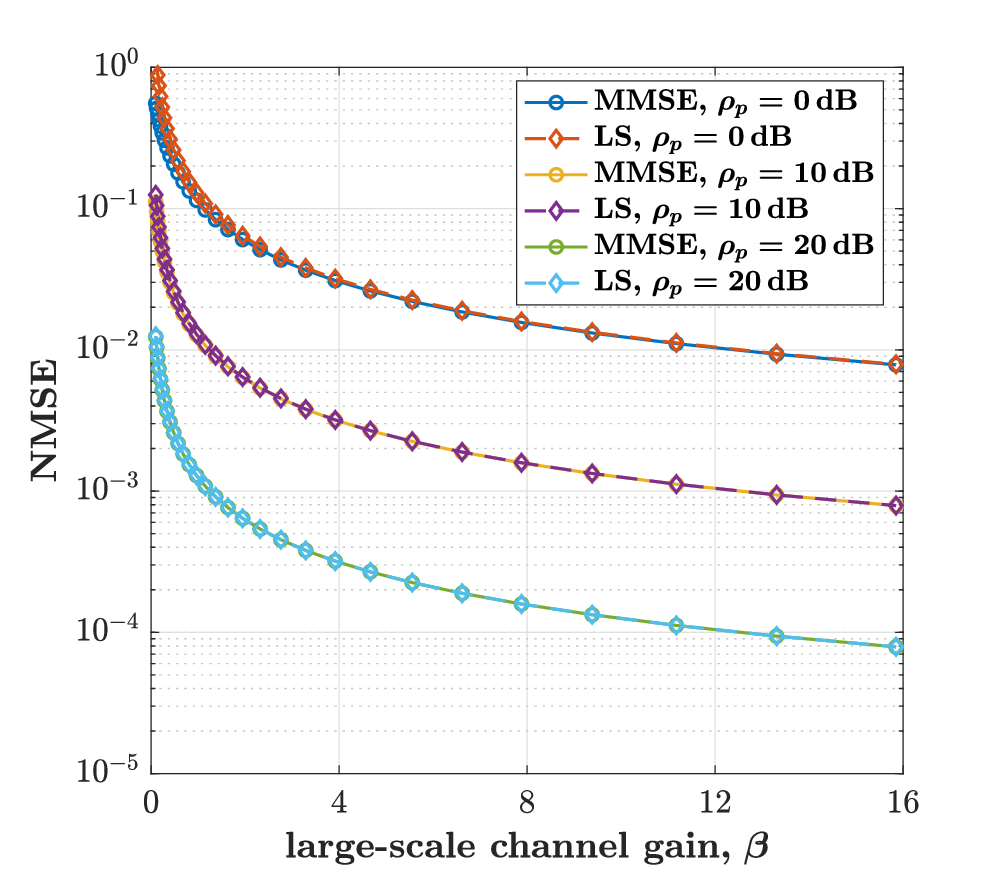}
\caption{NMSE versus routing gain $\beta$ for LS and MMSE estimators under different pilot SNR values.}\label{fig:nmse-beta}
\vspace{-5mm}
\end{figure}

Fig.~\ref{fig:nmse-beta} depicts the normalized mean square error (NMSE) of LS and MMSE channel estimators versus the large-scale channel gain $\beta$ under different pilot SNR levels $\rho_p \in \{0,10,20\}$ dB. As observed, the NMSE decreases monotonically with $\beta$. This behavior is expected since a larger $\beta$ increases the effective pilot observation power, thereby improving the SNR ratio of the training phase and reducing the estimation error. Increasing the pilot SNR $\rho_p$ further improves the NMSE across the entire $\beta$ range, as higher pilot power mitigates the impact of additive noise during channel estimation. Moreover, the MMSE estimator provides noticeable performance gains over LS in the low-$\beta$ and low-$\rho_p$ regimes, where the pilot observations are noise-limited. This improvement stems from the exploitation of prior second-order channel statistics in MMSE estimation, whereas LS relies solely on the received pilot signal. As either $\beta$ or $\rho_p$ increases, the pilot observation quality improves and the performance gap between MMSE and LS progressively diminishes. In high effective pilot SNR regimes, both estimators approach similar performance, reducing the relative benefit of statistical regularization.

\begin{figure}[]
	\centering
	\includegraphics[width=.9\columnwidth]{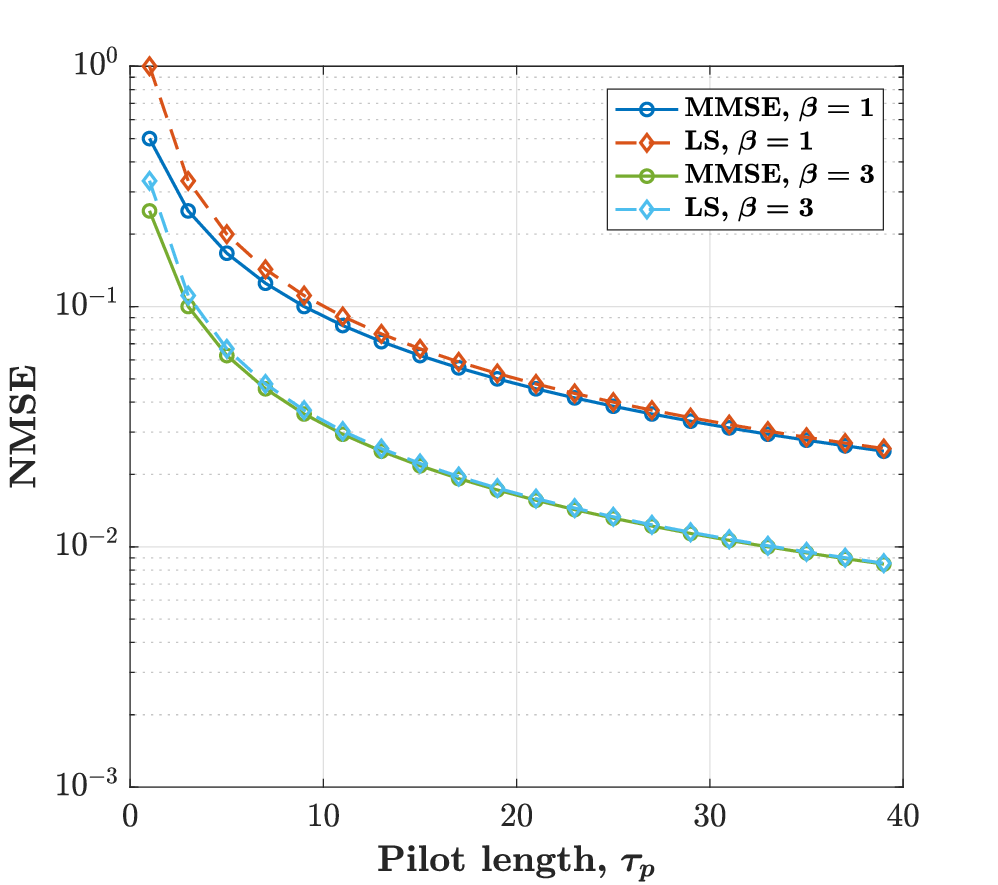}
	\caption{NMSE versus pilot length  $\tau_p$ at $\rho_p=0$ dB for different $\beta$.}\label{fig:nmse-pilot}
	\vspace{-2mm}
\end{figure}

Fig.~\ref{fig:nmse-pilot} illustrates the NMSE as a function of the pilot length $\tau_p$ at $\rho_p = 0$ dB for $\beta \in \{1,3\}$ under LS and MMSE estimation. The NMSE decreases monotonically with $\tau_p$ in all cases, as a longer pilot sequence increases the accumulated training energy $\tau_p \rho_p$ and therefore improves the effective observation SNR. For a fixed estimator, the case with $\beta = 3$ yields lower NMSE compared to $\beta = 1$, reflecting the fact that a stronger large-scale channel gain enhances the received pilot signal power and improves channel reconstruction reliability.  Moreover, the MMSE estimator provides lower NMSE than LS for both values of $\beta$, with the most pronounced difference observed at small $\tau_p$. In this regime, the limited training energy renders the estimation noise-limited, and the statistical regularization inherent in MMSE offers additional robustness. As $\tau_p$ increases, the performance gap progressively diminishes, since the improved observation quality reduces the relative impact of prior statistical information.

\begin{figure}[]
	\centering
	\includegraphics[width=\columnwidth]{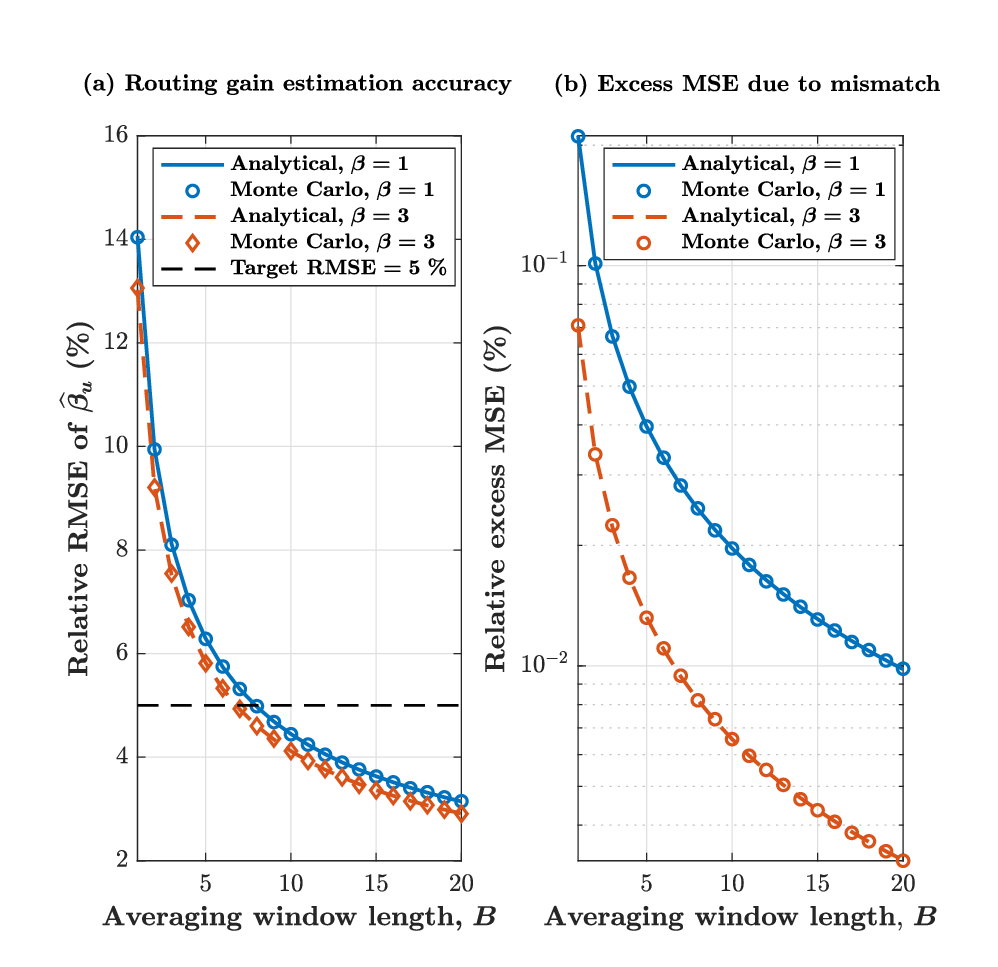}
\caption{	{\color{blue}Impact of the slow timescale averaging window length $B$ on: (a) the relative RMSE of the routing gain estimate and (b) the relative excess channel estimation MSE caused by mismatched MMSE estimation when $M=64$, $\tau_p=8$, and $\rho_p/\sigma^2=0$ dB. The horizontal dashed line in (a) indicates the target relative RMSE of $5\%$.}}\label{fig:bmis}
	\vspace{-2mm}
\end{figure}

{\color{blue}In Fig.~\ref{fig:bmis}, we evaluate the impact of the finite slow timescale averaging window on routing gain estimation and the resulting mismatched MMSE penalty. The close agreement between the analytical curves and the Monte Carlo markers validates the finite sample distribution and MSE analysis in Section~\ref{subsec:finite_B_mismatch}. As expected, increasing $B$ improves the routing gain estimation accuracy because a larger number of independent pilot energy observations is averaged. For $\beta_u=1$, the relative RMSE of $\widehat{\beta}_u$ decreases from $14.06\%$ at $B=1$ to $4.45\%$ at $B=10$. For $\beta_u=3$, the corresponding RMSE decreases from $13.02\%$ to $4.12\%$. According to \eqref{eq:B_selection_closed_form}, the target relative RMSE of $5\%$ is therefore satisfied with the minimum window lengths $B_u^\star=8$ and $B_u^\star=7$ for $\beta_u=1$ and $\beta_u=3$, respectively. The smaller required window for $\beta_u=3$ results from its higher effective pilot SNR. Fig.~\ref{fig:bmis}(b) further shows that the relative excess MSE caused by routing gain mismatch decreases rapidly with $B$. For $\beta_u=1$, the penalty decreases from approximately $0.211\%$ at $B=1$ to $0.0197\%$ at $B=10$, whereas for $\beta_u=3$ it decreases from approximately $0.0704\%$ to $0.00656\%$. These results indicate that the mismatched MMSE penalty can be effectively controlled through a moderate averaging window under the considered pilot SNR conditions. They also demonstrate that $B$ should be selected according to a prescribed estimation accuracy requirement rather than chosen arbitrarily. Besides, together with \eqref{eq:B_tracking_constraint}, these values imply that the $5\%$ accuracy target is feasible only if $B_r\geq8$ for $\beta_u=1$ and $B_r\geq7$ for $\beta_u=3$. Faster routing variations cannot support the prescribed accuracy within one routing stationarity interval.}

\begin{figure}[]
\centering
\includegraphics[width=.9\columnwidth]{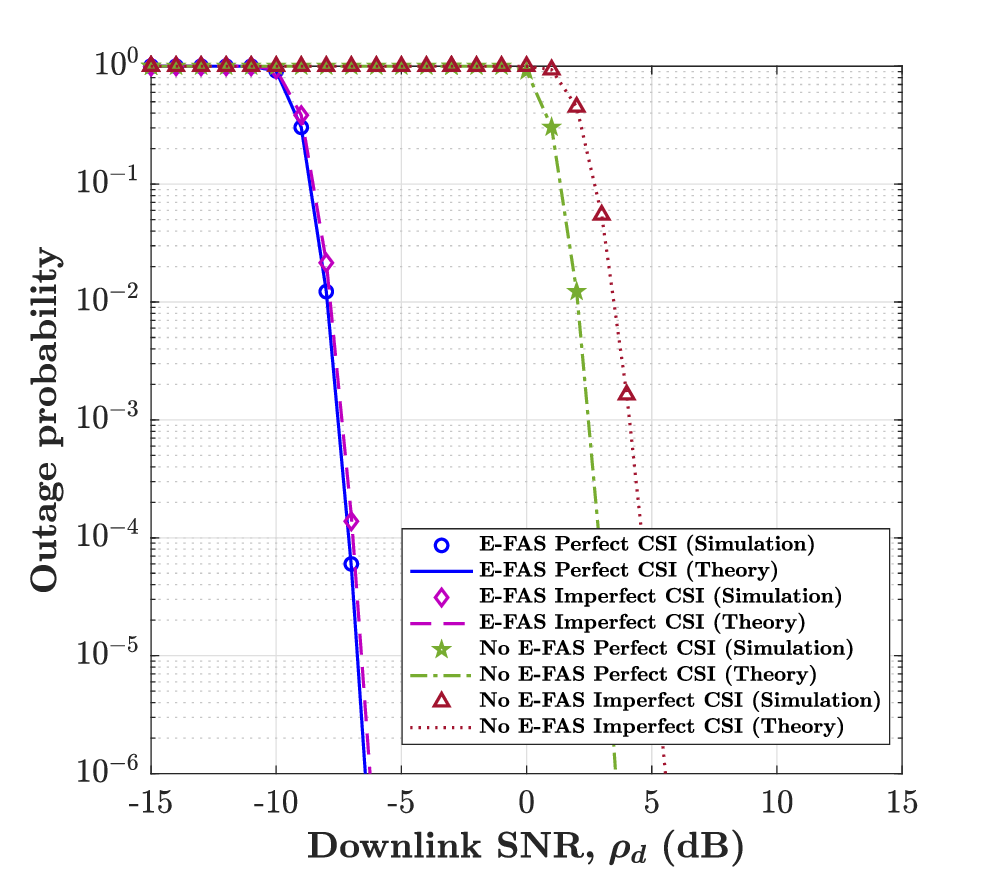}
\caption{Outage probability versus downlink SNR $\rho_d$ for $M=64$, $\tau_p=8$, and $\rho_p=0$ dB.}\label{fig:out-snr}
\vspace{-2mm}
\end{figure}

In Fig.~\ref{fig:out-snr}, we present the outage probability as a function of the downlink SNR $\rho_d$. The close agreement between the Monte Carlo simulations and the analytical curves confirms the correctness of the derived outage expressions. It is observed that the outage probability decreases monotonically with $\rho_d$ for all schemes, since increasing the transmit SNR enhances the received signal power and improves the likelihood that the instantaneous SINR exceeds the target threshold. A clear performance gap is observed between the E-FAS-assisted and conventional (No E-FAS) transmission. For a given outage level, the E-FAS scheme requires substantially lower $\rho_d$, which reflects the large-scale gain $\beta$ introduced by SW-assisted propagation. This gain effectively shifts the SINR distribution toward higher values, yielding a noticeable SNR advantage. As expected, perfect CSI provides an upper performance bound in both propagation scenarios. Channel estimation errors reduce the effective beamforming gain and introduce additional self-interference components arising from the mismatch between the true and estimated channels, resulting in a rightward shift of the outage curves. Under the considered parameter configuration, however, the performance gap between perfect and imperfect CSI remains moderate, particularly in the E-FAS case. This behavior is mainly due to the relatively high effective pilot SNR and the substantial array gain provided by $M=64$ antennas, which together enable accurate channel reconstruction and robust MRT beamforming.

\begin{figure}[]
\centering
\includegraphics[width=.9\columnwidth]{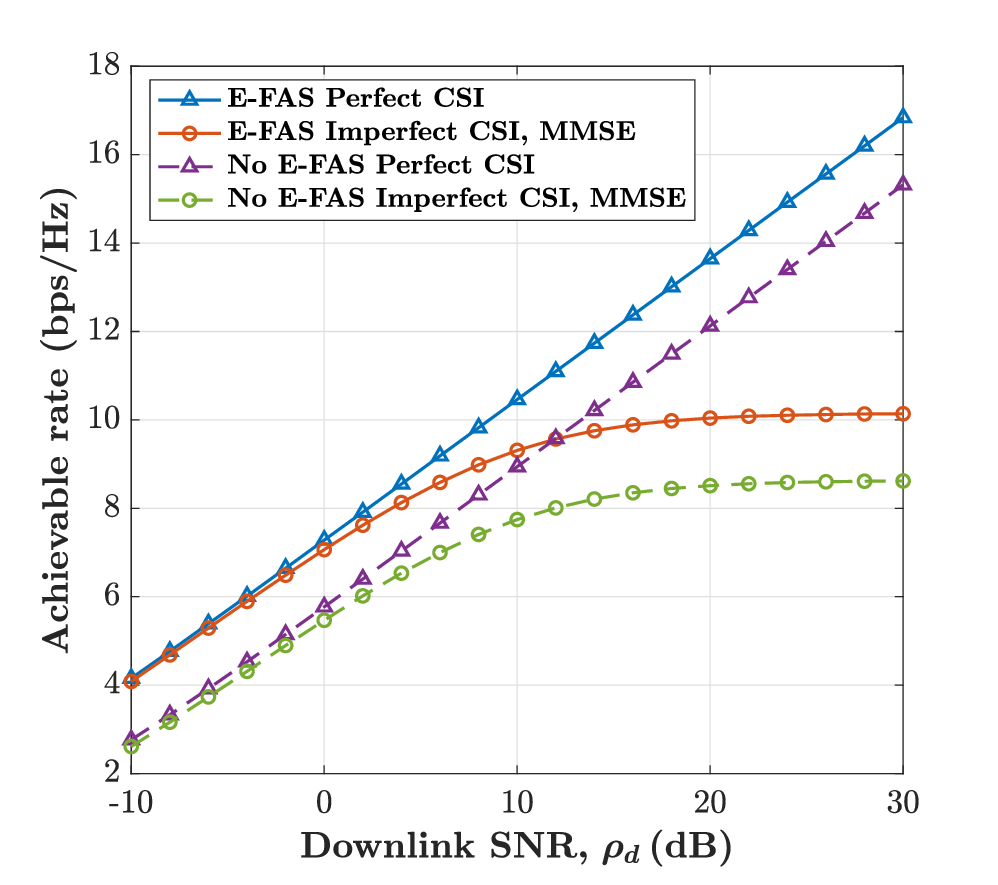}
\caption{Achievable single-user rate versus downlink SNR $\rho_d$ under MRT precoding with training overhead.}\label{fig:rate-snr}
\vspace{-2mm}
\end{figure}

Fig.~\ref{fig:rate-snr} shows the single-user rate versus the downlink SNR $\rho_d$, accounting for pilot overhead with $M=64$, $\tau_p=8$, and $T_c=200$. First, we see that the perfect CSI curves increase almost linearly with $\rho_d$ (in dB scale), reflecting the continuous growth of the effective received SNR due to coherent beamforming and array gain. In contrast, the imperfect CSI curves exhibit a clear saturation at high SNR. This behavior follows directly from the SINR expression under imperfect CSI, where the residual self-interference term $P\Omega_{\tilde h}$ scales linearly with the transmit power $P$. As $\rho_d$ increases, both the desired signal and the estimation error-induced interference grow proportionally, leading to a finite SINR ceiling and thus a rate saturation. {\color{blue} The resulting penalty is substantial. As quantified in Table~\ref{tab:csi_overestimation}, the perfect CSI assumption overestimates the E-FAS single-user rate by approximately $11.0\%$, $26.4\%$, and $39.8\%$ at downlink SNRs of $10$, $20$, and $30$ dB, respectively. Hence, the performance predicted under perfect CSI becomes progressively less representative as the downlink SNR increases while the pilot SNR remains fixed.} The E-FAS-assisted transmission consistently outperforms the conventional case across the entire SNR range. This gain originates from the larger large-scale channel coefficient $\beta$, which enhances both the beamforming gain and the quality of channel estimation. 

%Fig.~\ref{fig:rate-pilot} depicts the single-user rate versus the pilot length $\tau_p$ at $\rho_d=10$ dB. Under imperfect CSI, the rate exhibits a clear unimodal behavior. When $\tau_p$ is small, the channel estimation quality is poor due to insufficient training energy, resulting in a large estimation error variance and a reduced effective SINR. As $\tau_p$ increases, the improved channel estimation enhances the beamforming gain and increases the achievable rate. However, beyond a certain point, further increasing $\tau_p$ reduces the fraction of symbols available for data transmission through the pre-log penalty, causing the achievable rate to decrease. While this trade-off is inherent to pilot-based transmission, in the E-FAS architecture, the routing gain $\beta$ directly affects the effective pilot SNR and thus shifts the optimal training length. In contrast, under perfect CSI, no estimation error is present and increasing $\tau_p$ only reduces the available data transmission time. Consequently, the rate decreases monotonically with $\tau_p$ in the perfect CSI case. Furthermore, the E-FAS-assisted transmission consistently achieves higher rates than the conventional scenario due to the larger large-scale channel gain $\beta$, which enhances the effective received SNR. The MMSE estimator slightly outperforms the LS estimator across the considered range, as it yields a smaller estimation error variance and thus a higher effective SINR.

\begin{figure}[]
\centering
\includegraphics[width=.9\columnwidth]{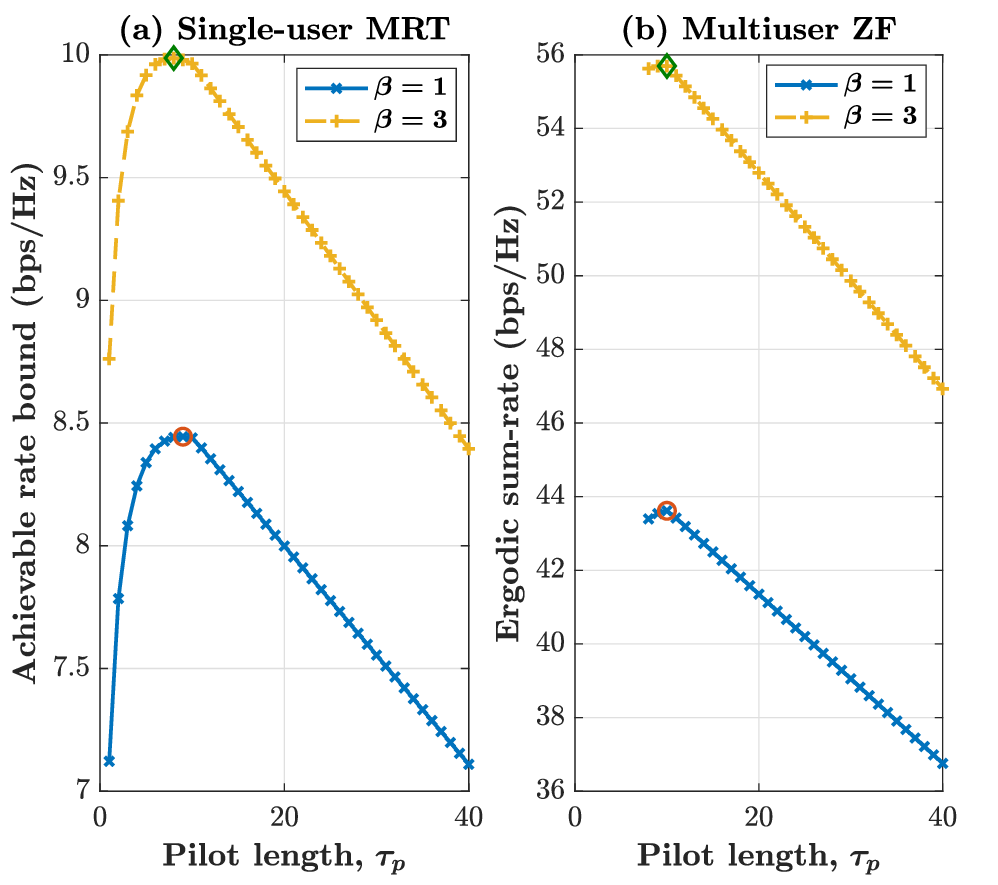}
%\caption{Achievable single-user rate versus pilot length $\tau_p$ at fixed SNR.}
\caption{{\color{blue}Effective rate versus pilot length $\tau_p$ under joint
	pilot power and pilot length allocation: (a) single-user MRT
	transmission and (b) multiuser ZF transmission with $K=8$.}}
	\label{fig:rate-pilot}
\vspace{-4mm}
\end{figure}

{\color{blue}Fig.~\ref{fig:rate-pilot} illustrates the joint pilot-power and pilot length trade-off. For a small $\tau_p$, the peak power constraint is active and increasing the pilot length increases the accumulated training energy, thereby improving the channel estimate. Once the pilot energy constraint becomes active, further increasing $\tau_p$ no longer increases the total training energy, while it continues to reduce the pre-log factor, producing a unique rate maximizing operating region. For single-user transmission, the optimal pilot lengths are $\tau_p^\star=9$ for $\beta=1$ and $\tau_p^\star=8$ for $\beta=3$, with $\rho_p^\star/\sigma^2=5$ dB in both cases. The resulting achievable rate bounds are approximately $8.44$ and $9.99$ bps/Hz, respectively. The shorter optimal training interval for $\beta=3$ follows from the higher effective pilot SNR provided by the E-FAS routing gain. For multiuser ZF transmission with $K=8$, the orthogonal pilot constraint limits the feasible pilot lengths to $\tau_p\geq8$. Under the considered resource budget, the optimal value is $\tau_p^\star=10$ for both $\beta=1$ and $\beta=3$, yielding sum-rates of approximately $43.62$ and $55.69$ bps/Hz, respectively. These results confirm that the allocation in \eqref{eq:optimal_pilot_length} balances estimation accuracy, pilot energy availability, and training overhead.}

\begin{figure}[]
\centering
\includegraphics[width=.9\columnwidth]{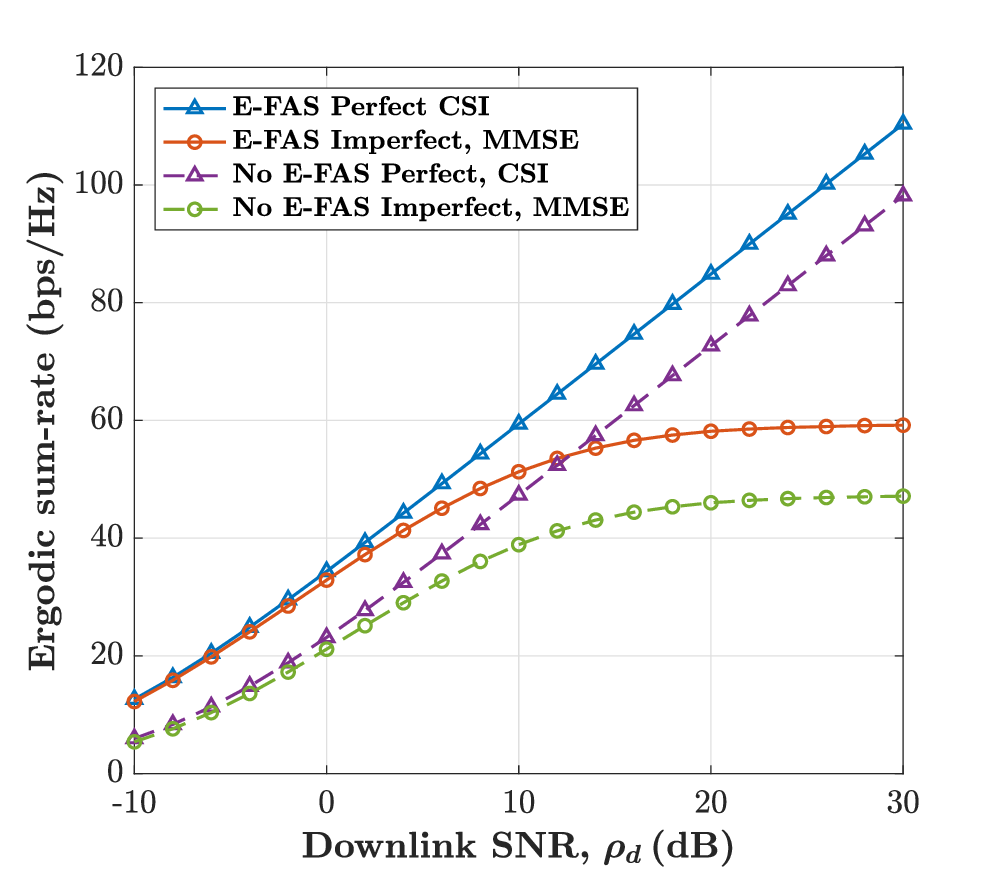}
\caption{Ergodic sum-rate versus downlink SNR $\rho_d$ for multiuser ZF transmission with training overhead.}\label{fig:sum-snr}
\vspace{-4mm}
\end{figure}

Fig.~\ref{fig:sum-snr} presents the multiuser ZF ergodic sum-rate versus the downlink SNR $\rho_d$ with pilot overhead, where $M=64$ and $K=8$. Under perfect CSI, the sum-rate increases almost linearly with $\rho_d$ (in dB scale), reflecting the full spatial multiplexing gain achieved by ZF precoding when inter-user interference is perfectly suppressed. The E-FAS-assisted system achieves a consistently higher sum-rate than the conventional transmission due to the enhanced large-scale channel gain, which effectively increases the received SNR for all users. In contrast, under imperfect CSI, the sum-rate exhibits a clear saturation behavior at high SNR. This phenomenon arises from residual multiuser interference caused by channel estimation errors. As the transmit power increases, both the desired signal and the interference terms scale proportionally with $P$, while the estimation error variance remains non-zero. As a result, the SINR converges to a finite limit, and the system becomes interference-limited rather than noise-limited. {\color{blue} This effect is even more pronounced in the multiuser setting. According to Table~\ref{tab:csi_overestimation}, neglecting channel estimation errors overestimates the E-FAS sum-rate by approximately $15.4\%$, $33.9\%$, and $48.5\%$ at downlink SNRs of $10$, $20$, and $30$ dB, respectively. The widening discrepancy results from the residual inter-user interference that remains after ZF precoding based on imperfect channel estimates.}  %Since both MMSE and LS produce estimation errors of comparable magnitude under the considered pilot configuration, their resulting sum-rate curves are closely aligned. 
Finally, the E-FAS-assisted transmission retains a noticeable advantage over the non-E-FAS case even under imperfect CSI, demonstrating that the large-scale routing gain not only enhances the desired signal power but also improves the robustness of spatial multiplexing.

\begin{figure}[t]
\centering
\includegraphics[width=.9\columnwidth]{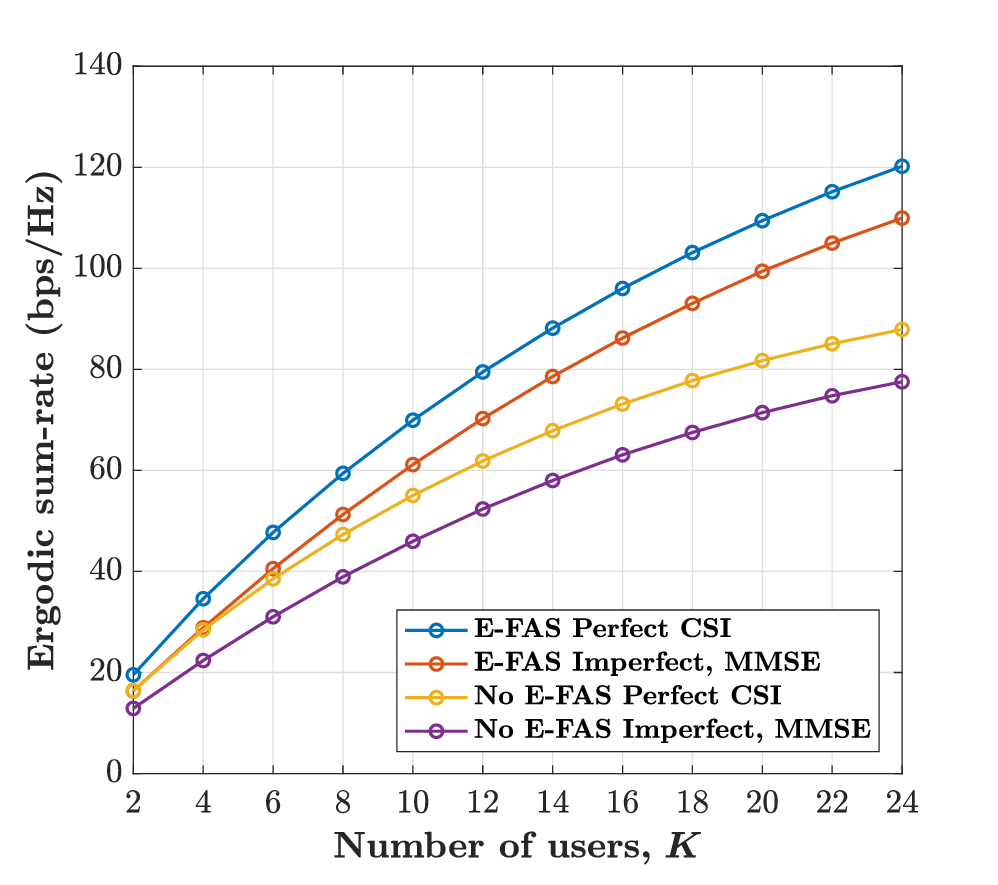}
\caption{Ergodic sum-rate versus number of users $K$ for ZF precoding at fixed SNR.}\label{fig:sum-user}
\vspace{-4mm}
\end{figure}

Fig.~\ref{fig:sum-user} illustrates the multiuser ZF ergodic sum-rate versus the number of served users $K$ at $\rho_d=10$ dB, with $M=64$  and orthogonal pilot allocation $\tau_p=K$. We see that the sum-rate increases with $K$ over the shown range. This trend reflects the spatial multiplexing gain of the system: serving more users allows the BS to exploit additional spatial DoF, thereby increasing the total throughput. Since $M=64$ is significantly larger than the values of $K$, the ZF precoder can effectively separate users, and the multiplexing gain dominates. However, increasing $K$ also increases the pilot overhead due to the orthogonal pilot requirement $\tau_p=K$, which reduces the effective pre-log factor $(1-\tau_p/T_c)$. Consequently, although the sum-rate grows with $K$ in the considered range, the growth rate gradually decreases, indicating the emerging impact of training overhead. For sufficiently large $K$, the pre-log penalty would eventually dominate and limit further gains. It is further observed that the E-FAS-assisted transmission consistently achieves higher sum-rate than the conventional scenario without E-FAS. This gain stems from the enhanced large-scale channel gain, which improves the effective SINR of all users and strengthens the spatial multiplexing performance. Additionally, it is observed that the imperfect CSI introduces a noticeable performance degradation compared to perfect CSI due to residual multiuser interference. Nevertheless, the performance gap remains bounded, indicating that the proposed system retains substantial multiplexing benefits even under practical channel estimation constraints.

\begin{figure}[t]
	\centering
	\includegraphics[width=\columnwidth]{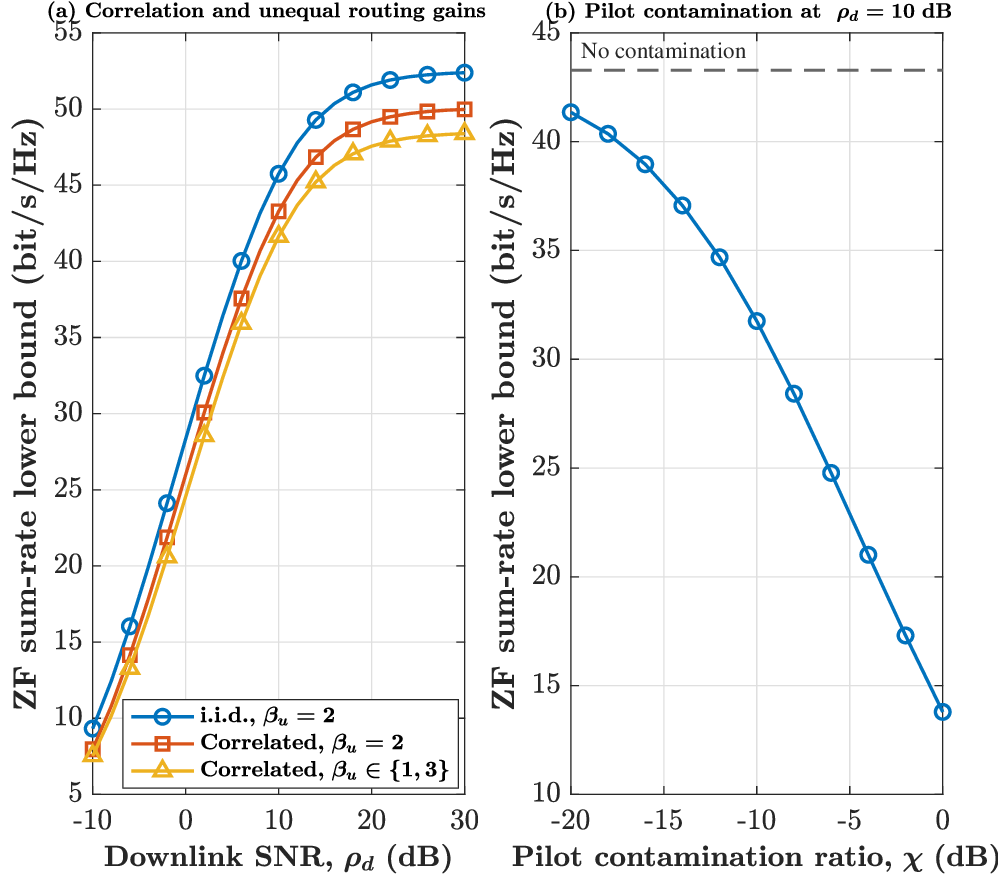}
\caption{{\color{blue}ZF sum-rate robustness beyond the i.i.d. orthogonal pilot
	model: (a) spatial correlation and unequal normalized routing gains;
	(b) pilot-contamination sensitivity at $\rho_d=10$ dB.}}\label{fig:correl}
	\vspace{-2mm}
\end{figure}

{\color{blue}
Fig.~\ref{fig:correl} evaluates the generalized channel and pilot models through Monte Carlo simulations, rather than validating the i.i.d.~closed forms. In panel (a), spatial correlation is modeled as $[\mathbf R_u]_{m,n}=0.7^{|m-n|}$ with $\operatorname{tr}(\mathbf R_u)=M$ and $\overline{\mathbf h}_u=\mathbf 0$. The equal-routing case uses $\beta_u=2$ for all users, whereas the unequal-routing case assigns $\beta_u=1$ and $\beta_u=3$ to four users each, preserving the same average routing gain. The covariance-dependent ZF SINR 
is evaluated without the scalar error covariance simplification. Spatial correlation and routing gain imbalance both reduce the sum-rate, while preserving the fixed training saturation behavior. Panel (b) considers one external user reusing each in-cell pilot, with contamination ratio $\chi=q_{c,u}\beta_{c,u}/(q\beta_u)$, $\beta_u=\beta_{c,u}=2$, and $\rho_d=10$ dB. Increasing $\chi$ degrades the channel estimate and increases residual leakage after ZF precoding, thereby reducing the sum-rate lower bound. This experiment isolates training contamination; data phase inter-cell interference is not included.}

\vspace{0mm}

\begin{figure}[t]
	\centering
	\includegraphics[width=.8\columnwidth]{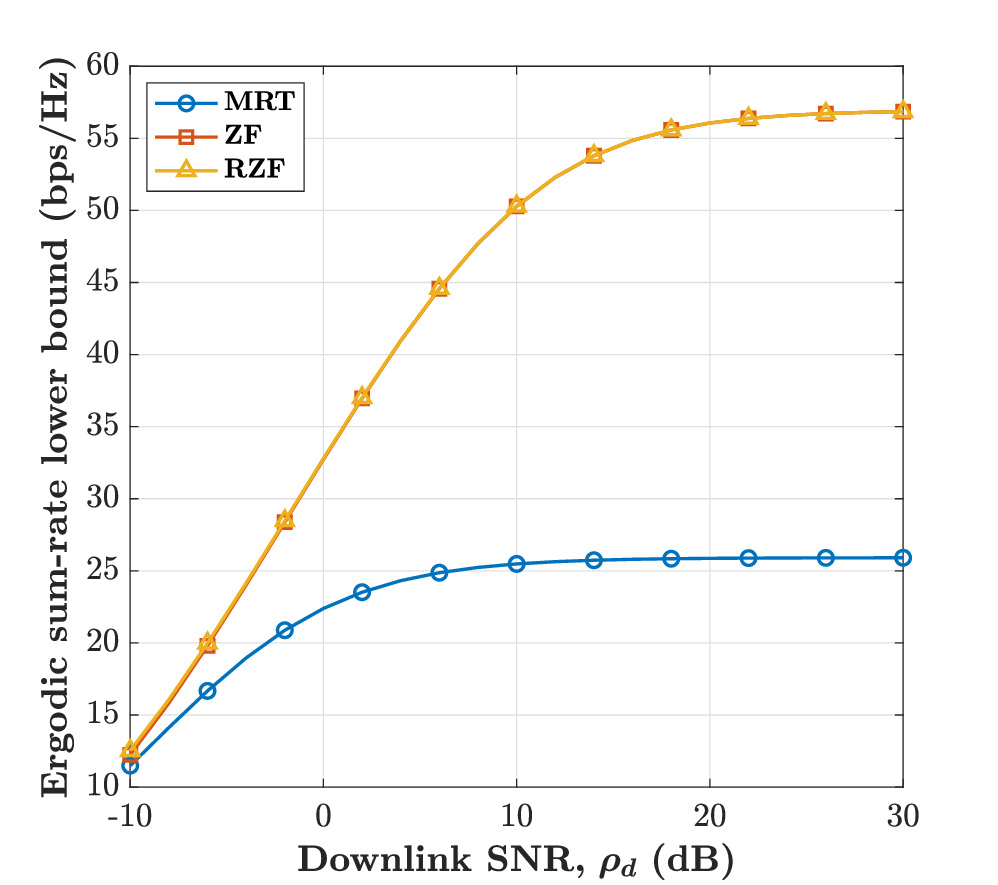}
	\caption{\color{blue}Ergodic sum-rate comparison of MRT, ZF, and RZF under imperfect MMSE CSI.}\label{fig:prc}
%	\vspace{-4mm}
\end{figure}

{\color{blue}Fig.~\ref{fig:prc} 
compares MRT, ZF, and  regularized ZF (RZF) under identical pilot resources, equal power allocation, and imperfect MMSE CSI, where RZF uses regularization $\xi=K\sigma^2/P$. RZF closely tracks ZF and converges to it as $\xi$ vanishes. At moderate and high SNR, ZF and RZF outperform MRT because MRT retains estimated-channel inter-user interference. Under fixed pilot resources, all schemes eventually
saturate, with MRT exhibiting the lowest ceiling.}

%\vspace{-4mm}

\section{Conclusion}\label{sec:conc}
This paper investigated the impact of imperfect CSI on E-FAS-assisted downlink transmission. We considered a practical pilot-based estimation framework and developed a tractable MMSE channel estimation model for the equivalent E-FAS channel. Closed-form expressions were derived for the statistics of the estimated channel and the estimation error. For the single-user case, analytical expressions for outage probability upper bound and an ergodic rate lower bound were obtained under imperfect CSI, {\color{blue}revealing a fixed training high-SNR saturation caused by residual estimation errors.} In the multiuser case, ZF precoding based on imperfect CSI estimates was studied, and the resulting residual interference-limited SINR and ergodic sum-rate behavior were
characterized. 

{\color{blue} Under the configuration with $M=64$, $\beta=3$, $\tau_p=8$, and a pilot SNR of $0$ dB, the perfect CSI assumption overestimates the single-user rate and the multiuser ZF sum-rate by approximately $39.8\%$ and $48.5\%$, respectively, at a downlink SNR of $30$ dB. These results confirmed that practical channel acquisition materially affects the performance predicted for E-FAS, particularly in the high-SNR regime under fixed pilot resources.} {\color{blue}The finite window analysis further demonstrated that routing-gain uncertainty and its mismatched MMSE penalty decrease rapidly with the averaging-window length. Under the nominal setting with $M=64$, $\tau_p=8$, and a pilot SNR of $0$ dB, a target relative routing-gain RMSE of $5\%$ requires $B=8$ blocks for $\beta_u=1$ and $B=7$ blocks for $\beta_u=3$. At $B=10$, the corresponding relative excess MSE penalties are approximately $0.0197\%$ and $0.00656\%$, respectively.}

{\color{blue}Furthermore, under fixed pilot resources, non-zero estimation errors impose a high-SNR performance ceiling, whereas scaling the pilot SNR with the downlink SNR can eliminate this ceiling. The E-FAS power gain advantage refers to the normalized comparison in which $\beta=1$ represents the conventional link baseline and $\beta>1$ represents a relative E-FAS routing enhancement. Under identical transmit power and pilot resource conditions, a larger $\beta$ increases both the received data signal power and the effective pilot SNR, resulting in improved outage and rate bounds despite residual CSI errors.}

{\color{blue}Future work will consider spatially correlated and mobility-aware channel estimation for E-FAS, joint optimization of routing, pilot resources, and precoding, and validation through electromagnetic simulations and E-FAS prototypes.}

%\vspace{-3mm}
\appendices
\section{Closed-Form Evaluation of the Achievable Rate}\label{app:rate_integral}
Consider the integral in \eqref{eq:int} as\vspace{-0.2cm}
\begin{align}
	I \triangleq
	\int_{0}^{\infty}
	\ln\!\left(
	1+\frac{P x}{P\Omega_{\tilde{h}}+\sigma^2}
	\right)
	\frac{x^{M-1}}
	{\Gamma(M)\Omega_{\hat{h}}^{M}}
	\mathrm{e}^{-\frac{x}{\Omega_{\hat{h}}}}
	\,dx,
	\label{eq:I_def}
\end{align}
where $M\in\{1,2,\ldots\}$, $\Omega_{\hat{h}}>0$, and $P\Omega_{\tilde{h}}+\sigma^2>0$.
By defining the positive constants as
$
	a \triangleq \frac{P}{P\Omega_{\tilde{h}}+\sigma^2}
$, $
	z \triangleq \frac{1}{a\Omega_{\hat{h}}}
	= \frac{P\Omega_{\tilde{h}}+\sigma^2}{P\,\Omega_{\hat{h}}}
%	\label{eq:a_z_def}
$, and the change of variables \(x=\Omega_{\hat{h}}t\), \eqref{eq:I_def} becomes
\begin{align}
	I
	=
	\frac{1}{\Gamma(M)}
	\int_{0}^{\infty}
	\ln\!\left(1+\frac{t}{z}\right)
	t^{M-1}e^{-t}\,dt.
	\label{eq:I_scaled}
\end{align}
Now, let $F_M(z)\triangleq I$. 
%\begin{align}
%	F_M(z) \triangleq
%	\frac{1}{\Gamma(M)}
%	\int_{0}^{\infty}
%	\ln\!\left(1+\frac{t}{z}\right)
%	t^{M-1}e^{-t}\,dt.
%	\label{eq:F_def}
%\end{align}
By differentiating $F_M(z)$ under the integral sign, we have
\begin{align}
	\frac{dF_M(z)}{dz}
	&=
	\frac{1}{\Gamma(M)}
	\int_{0}^{\infty}
	\frac{\partial}{\partial z}
	\ln\!\left(1+\frac{t}{z}\right)
	t^{M-1}e^{-t}\,dt
	\nonumber\\
%	&=
%	-\frac{1}{\Gamma(M)}
%	\int_{0}^{\infty}
%	\frac{t}{z(z+t)}\,t^{M-1}e^{-t}\,dt
%	\nonumber\\
	&=
	-\frac{1}{z\Gamma(M)}
	\int_{0}^{\infty}
	\frac{t^{M}}{z+t}\,e^{-t}\,dt.
	\label{eq:Fprime_start}
\end{align}
Using the identity \(\frac{1}{z+t}=\int_{0}^{\infty}e^{-(z+t)u}\,du\) for \(z>0\),
\eqref{eq:Fprime_start} becomes
\begin{align}
	\frac{dF_M(z)}{dz}
%	&=
%	-\frac{1}{z\Gamma(M)}
%	\int_{0}^{\infty}
%	\left(
%	\int_{0}^{\infty} t^M e^{-t} e^{-(z+t)u}\,dt
%	\right) du
%	\nonumber\\
	=
	-\frac{1}{z\Gamma(M)}
	\int_{0}^{\infty}
	e^{-zu}
	\left(
	\int_{0}^{\infty} t^M e^{-(1+u)t}\,dt
	\right) du.
	\label{eq:Fprime_swap}
\end{align}
Since \(\int_{0}^{\infty} t^M e^{-(1+u)t}\,dt=\Gamma(M+1)(1+u)^{-(M+1)}\),
\eqref{eq:Fprime_swap} yields
\begin{align}
	\frac{dF_M(z)}{dz} =
%	&=
%	-\frac{\Gamma(M+1)}{z\Gamma(M)}
%	\int_{0}^{\infty}
%	\frac{e^{-zu}}{(1+u)^{M+1}}\,du
%	\nonumber\\
%	&=
	-\frac{M}{z}
	\int_{0}^{\infty}
	\frac{e^{-zu}}{(1+u)^{M+1}}\,du.
	\label{eq:Fprime_u}
\end{align}
With the substitution \(s=1+u\), we obtain
\begin{align}
\hspace{-0.3cm}	\frac{dF_M(z)}{dz}
	=
	-\frac{M}{z}e^{z}
	\int_{1}^{\infty}
	\frac{e^{-zs}}{s^{M+1}}\,ds
	=
	-\frac{M}{z}e^{z}\mathrm{E}_{M+1}(z),
	\label{eq:Fprime_E}
\end{align}
where the generalized exponential integral is defined as
$
	\mathrm{E}_n(z)\triangleq \int_{1}^{\infty}\frac{e^{-zt}}{t^n}\,dt,
$ with $ z>0,\ n\ge 1.
%	\label{eq:En_def}
$
Using the standard recurrence
$
	n\mathrm{E}_{n+1}(z)=e^{-z}-z\mathrm{E}_n(z)$ with
	$ n\ge 1$,
%	\label{eq:En_recur}
\eqref{eq:Fprime_E} simplifies to
\begin{align}
	\frac{dF_M(z)}{dz}
	=
	-\frac{1}{z}+e^{z}\mathrm{E}_M(z).
	\label{eq:Fprime_simple}
\end{align}
Now, we define
$
	G_M(z)\triangleq e^{z}\sum_{k=1}^{M}\mathrm{E}_k(z)
%	\label{eq:G_def}
$.
Using the derivative identity \(\frac{d}{dz}\mathrm{E}_k(z)=-\mathrm{E}_{k-1}(z)\) for \(k\ge 2\) and
\(\frac{d}{dz}\mathrm{E}_1(z)=-e^{-z}/z\), it follows that
\begin{align}
	\frac{dG_M(z)}{dz}
	=
	-\frac{1}{z}+e^{z}\mathrm{E}_M(z),
	\label{eq:Gprime}
\end{align}
 which matches \eqref{eq:Fprime_simple}. Therefore, $F_M(z)=G_M(z)+C$ for some constant $C$.
Since $F_M(z)\to 0$ as $z\to\infty$ (equivalently $a\to 0$), and
$e^{z}\mathrm{E}_k(z)\to 0$ as $z\to\infty$, we have $C=0$. Hence,
$
	I
	=
	F_M(z)
	=
	e^{z}\sum_{k=1}^{M}\mathrm{E}_k(z)$ with 
	$z=\frac{P\Omega_{\tilde{h}}+\sigma^2}{P\,\Omega_{\hat{h}}}
%	\label{eq:I_closed_form}
$.

%\vspace{-4mm}

\bibliographystyle{IEEEtran}

\end{document}